\documentclass[a4paper]{article}
\usepackage{graphicx}

\newcommand{\ket}[1]{ \left| \: #1 \right>}

\usepackage{geometry}         
\usepackage[francais]{babel}  

\title{Sym\'etrie et th\'eorie des groupes
\`a travers la physique}           
\author{Jacques Villain}

\begin{document}

\maketitle                  

\begin{abstract}
  Les propri\'et\'es de la mati\`ere sont li\'ees, comme l'avait remarqu\'e Pierre Curie en 1884, au type de sym\'etrie qui y r\`egne. La th\'eorie des groupes est un outil syst\'ematique, mais pas toujours commode, pour exploiter  
cette sym\'etrie, notamment quand il faut trouver les vecteurs propres et valeurs propres 
d'un op\'erateur. 
Certaines propri\'et\'es (pouvoir rotatoire optique, pi\'ezo\'electricit\'e...) 
sont interdites dans des cristaux ou 
mol\'ecules de haute sym\'etrie. Quelques th\'eor\`emes (Noether, Goldstone...) \'etablissent des
 relations g\'en\'erales entre les propri\'et\'es physiques et la sym\'etrie. On passe en revue quelques applications de la th\'eorie des groupes 
\`a la physique de la mati\`ere condens\'ee, \`a la physique des particules \'el\'ementaires, \`a la m\'ecanique quantique,
\`a l'\'electromagn\'etisme. La th\'eorie des groupes n'est pas seulement un outil, mais aussi une belle construction 
qui permet de mieux comprendre la nature.  
\end{abstract}
%
\section{Sym\'etrie et propri\'et\'es physiques}

\begin{figure}[t]
\centering
 \includegraphics*[width=70mm, ]{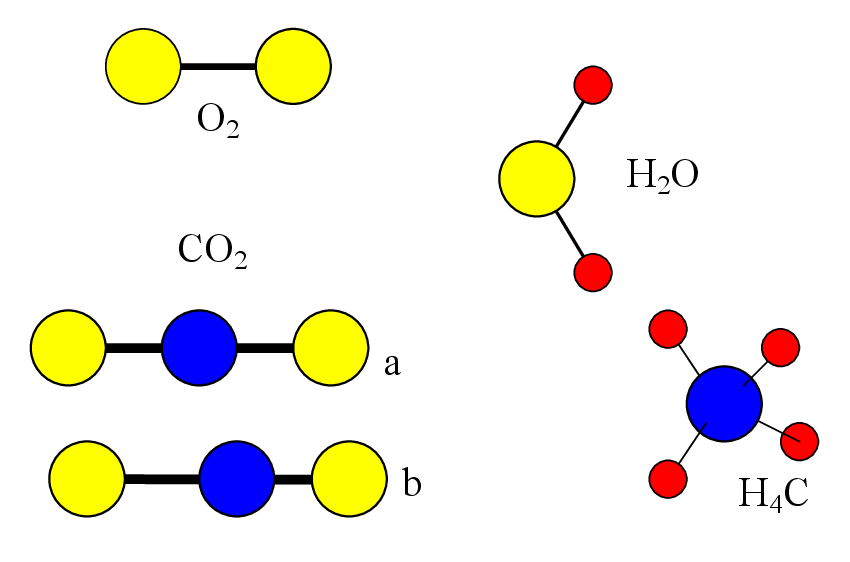}
\caption{Sym\'etrie de diff\'erentes mol\'ecules. L'oxyg\`ene est centrosym\'etrique et conserve sa sym\'etrie quand il vibre. Il en r\'esulte qu'il ne peut ni absorber ni \'emettre du rayonnement infrarouge. L'eau, le dioxyde de carbone et le m\'ethane ont  un moment \'electrique variable quand ils vibrent, peuvent absorber et \'emettre du rayonnement infrarouge. Le dioxyde de carbone a, par sym\'etrie, un moment \'electrique moyen nul (a) mais acquiert un moment non nul quand il vibre (b). 
Le m\'ethane a une propri\'et\'e analogue.
}
\label{fig1}
\end{figure}

Notre vie est possible gr\^ace aux propri\'et\'es physique des mol\'ecules qui nous entourent : 
O$_2$, N$_2$, H$_2$O.... Or ces propri\'et\'es physiques d\'ependent beaucoup de la sym\'etrie. 
Ainsi les mol\'ecules O$_2$ et N$_2$ sont sym\'etriques par rapport \`a leur centre de gravit\'e (Fig. \ref{fig1}). Cela implique que leur moment \'electrique est nul. 
Cette propri\'et\'e subsiste quand la mol\'ecule vibre. Il en r\'esulte que l'air sec n'absorbe pas le rayonnement infrarouge \'emis par la terre\footnote{Cependant les mol\'ecules d'oxyg\`ene et d'azote peuvent acqu\'erir un moment 
\'electrique si les \'electrons se d\'eplacent par rapport aux noyaux, mais cela correspond \`a des fr\'equences de vibration \'elev\'ees qui sont celles de la lumi\`ere visible. L'oxyg\`ene et l'azote absorbent donc une partie de la 
lumi\`ere visible venant du soleil.}. L'effet de serre est donc d\^u uniquement \`a l'eau et \`a d'autres mol\'ecules moins sym\'etriques. Les mol\'ecules les plus abondantes dans l'air (O$_2$ et N$_2$) ne contribuent pas \`a l'effet de serre. C'est un r\'esultat essentiel, et pourtant sa d\'emonstration est \'el\'ementaire. 

Nous donnerons un autre exemple qui concerne un mat\'eriau \`a la mode actuellement, le graph\`ene\cite{Wunsch}. 
Sa structure \'electronique 
a un propri\'et\'e remarquable: la bande de conduction et la bande de valence sont situ\'ees l'une au dessus de l'autre 
mais se touchent en un nombre fini de points de l'espace r\'eciproque, \`a la diff\'erence de ce qui se passe soit dans un m\'etal, soit dans un isolant ou un semi-conducteur. Nous montrons dans 
l'appendice \ref{secA1}  comment cela r\'esulte de la 
sym\'etrie. 

Ce sont l\`a des exemples  de la simplification 
souvent apport\'ee par la sym\'etrie dans 
les probl\`emes de physique. Il existe des probl\`emes o\`u la sym\'etrie est plus compliqu\'ee, mais 
n\'eanmoins permet de faciliter l'analyse des propri\'et\'es physiques. 
Il est alors souvent utile d'utiliser la notion de groupe 
et la th\'eorie des groupes. 

Le pr\'esent texte est la version fran\c caise d'un cours introductif pr\'esent\'e \`a une \'ecole
sur la sym\'etrie en physique de la mati\`ere condens\'ee, tenue en mai 2009. Nous ferons de nombreuses r\'ef\'erences
aux autres cours de cette \'ecole, qui seront publi\'es en 2010 au Journal de Physique IV. 
Cette publication aura l'avantage de r\'eunir de fa\c con compacte la plupart  des \'el\'ements n\'ecessaires \`a un approfondissement,
mais cet approfondissement peut aussi se trouver dans les autres r\'ef\'erences dispers\'ees auxquelles nous renvoyons.

Nous essaierons d'\'eviter un expos\'e trop superficiel, 
tout en \'evitant des complications techniques excessives. Ce juste milieu est difficile \`a trouver,
et nous comptons sur l'indulgence de nos lecteurs.

\section{D\'efinition du concept de groupe}

Un groupe est un ensemble d'\'el\'ements $g_i$ muni d'une r\`egle d'association interne (que nous appellerons en g\'en\'eral multiplication) qui associe \`a toute paire $g_i$, $g_j$ un autre \'el\'ement de l'ensemble que nous appellerons  produit et que nous noterons $g_i . g_j$ ou $g_i . g_j$. Il doit y avoir un \'el\'ement neutre 1 tels que $g_i . 1= 1 . g_i = g_i$. Enfin chaque \'el\'ement $g_i$ doit avoir un inverse $g_i^{-1}$ tel que 
$g_i .g_i^{-1} = g_i^{-1}.g_i=1$.
Un groupe peut \^etre fini ou infini. Il est dit commutatif ou ab\'elien si $g_i . g_j=g_j . g_i$.
Les transformations qui laissent un syst\`eme invariant forment un groupe. L'\'el\'ement neutre est alors constitu\'e par la "transformation identique" ou "identit\'e" 1, qui transforme un objet en lui-m\^eme. Par exemple, la sym\'etrie $I$ par rapport \`a un point forme avec l'identit\'e un groupe de 2 \'el\'ements. C'est ce groupe que nous avons 
consid\'er\'e au paragraphe pr\'ec\'edent, sans prononcer le mot, car pour un probl\`eme aussi simple la notion de groupe et la th\'eorie des groupes sont inutiles. 

Le groupe de transformations qui laissent un objet invariant est appel\'e la sym\'etrie de cet objet. Ce mot "sym\'etrie" a donc une signification plus large que celle qu'elle a dans la vie courante. 

\section{Quelques groupes qui servent au physicien}
\label{sec3}

Il est utile de faire une liste de groupes que les physiciens utilisent souvent. Le lecteur est 
invit\'e \`a survoler rapidement cette \'enum\'eration pour y revenir quand ce sera n\'ecessaire.

1) Le groupe SO3 ou SO(3) des rotations de l'espace \`a 3 dimensions autour d'un point donn\'e. Dans le traitement non relativiste de l'atome d'hydrog\`ene, le hamiltonien est invariant par SO3 et cela facilite la recherche des fonctions d'onde. On peut d\'efinir plus g\'en\'eralement  le groupe SO($n$). 
2) Le groupe O3 ou O(3) des rotations propres et impropres 3-D. Une rotation impropre est le produit d'une rotation par une sym\'etrie par rapport \`a un point (ou \`a un plan).

3) Le groupe des translations 3-D. Il intervient dans la propagation d'une particule libre non relativiste.

4) Le groupe des rotations et translations 3-D.
 
5) Groupe des translations qui conservent un cristal (sous-groupe de (3)). 

6) Groupe des rotations propres et impropres qui conservent un cristal (sous-groupe de (4)). C'est le groupe "d'espace" du cristal. 

7) Groupe des rotations propres et impropres autour d'un axe, passant par un point donn\'e, qui conservent un cristal (sous-groupe fini de (3)) : groupe "ponctuel". 

8) Groupe de Lorentz, groupe de Poincar\'e. Ce sont des groupes de transformations adapt\'ees \`a la relativit\'e. En relativit\'e, on ne peut pas traiter le temps $t$ ind\'ependamment des coordonn\'ees d'espace $x, y, z$. Il faut donc faire intervenir des transformations qui agissent sur les 4 coordonn\'ees 
$x, y, z, t$. Par exemple celle qui transforme $(x, y, z, t)$ en
\begin{equation}
x'=(x- vt)/(1-\beta ^2) ^{1/2}\;\;; \;\; t'=(t- vx/c)/(1-\beta ^2) ^{1/2}\;\;; \;\;y'=y \;\;; \;\;z'=z 
\label{Lorentz}
\end{equation}
o\`u $\beta =v/c$. Cette transformation d\'ecrit le passage d'un rep\`ere particulier \`a un autre rep\`ere
en translation uniforme \`a vitesse $v$ dans la direction $x$ par rapport au premier. 
On remarque qu'elle conserve la quantit\'e 
\begin{equation}
ds^2=dx^2+dy^2+dz^2-c^2dt^2 
\label{Minko}
\end{equation}
alors qu'une rotation ou une translation tridimensionnelle conserve
\begin{equation}
ds^2=dx^2+dy^2+dz^2 
\label{Euclid}
\end{equation}
Toute transformation qui conserve (2) s'appelle une transformation de Lorentz. 
Les transformations de Lorentz constituent le groupe de Poincar\'e.
On appelle groupe de Lorentz l'ensemble des transformations de Lorentz qui conservent l'origine (0,0,0,0). 

9) SU2, SU3... SU$n$...= groupe des matrices $n\times n$ unitaires de d\'eterminant 1. On \'ecrit aussi SU($n$).

10) U($n$) est le groupe des matrices unitaires $n\times n$. Par exemple, U(1) est form\'e par les nombres 
$\exp(i\varphi )$.
Les groupes continus tels que O($n$) font partie d'une classe plus g\'en\'erale dont les \'el\'ements sont appel\'es groupes de Lie. 

En r\'esum\'e : la notion de groupe est partout pr\'esente en physique. La question est maintenant : \`a quoi peut-elle servir? On va donner des exemples.

\section{Cristallographie}
\label{sec4}

Un cristal est  la forme thermodynamiquement stable de la plupart des solides \`a basse temp\'eratures. 
D'o\`u l'importance des cristaux. 

Comme on l'a dit plus haut, la sym\'etrie d'un cristal est d\'efinie par son groupe d'espace, dont chaque \'el\'ement est le produit d'une translation par une rotation ${\mathcal R}$ autour 
d'un point O donn\'e. Si on ne consid\`ere que les rotations ${\mathcal R}$, 
elles forment aussi un groupe (fini) appel\'e groupe ponctuel. Pour des d\'etails,
on se pourra se reporter aux articles de  Grenier (2009) et  d'Aroyo (2009). 
Certaines propri\'et\'es ne d\'ependent que du groupe ponctuel. 
D'autre part un cristal \`a un seul atome par maille (comme le fer et beaucoup d'autres corps simples) 
s'appelle un r\'eseau de Bravais. 
Il existe 32 groupes ponctuels et 230 groupes d'espace, comme l'ont d\'emontr\'e ind\'ependamment Sch\"onflies et Fedorov en 1890. Il existe 14 r\'eseaux de Bravais, comme l'a montr\'e Bravais en 1848. Ces nombres r\'esultent de la solution d'un pur probl\`eme de math\'ematique, comparable \`a l'\'enum\'eration descriptive des 5 poly\`edres r\'eguliers, 
donn\'ee par Euclide\footnote{Pour donner une id\'ee de la d\'emonstration, montrons que toute rotation d'angle 
$\varphi <\pi /3$ est interdite. Si une telle rotation laissait un r\'eseau de Bravais invariant, ce r\'eseau contiendrait 3 points ABC avec AB=BC et l'angle ABC \'egal \`a $\varphi <\pi /3$ . La longueur AC serait donc 
inf\'erieure \`a AB et BC. Mais  le 
r\'eseau devrait alors aussi contenir un point D avec AD=AC et l'angle DAC \'egal \`a $\varphi <\pi /3$. La longueur DC serait donc inf\'erieure \`a AD et AC. Par r\'ecurrence on peut donc trouver des longueurs aussi petites que l'on veut, ce qui est impossible puisque la distance interatomique ne peut \^etre inf\'erieure, disons, \`a 0,1 microm\`etre}. 
Le d\'enombrement et la description des groupes d'espace et ponctuels n'est pas une application de la th\'eorie des groupes, mais plut\^ot une partie de cette th\'eorie.

A quoi servent groupes d'espace et ponctuels? Certaines propri\'et\'es physiques existent ou n'existent pas selon la sym\'etrie du cristal, c'est-\`a-dire selon son groupe d'espace ou son groupe ponctuel. 
Un exemple est la pyro\'electricit\'e, caract\'eris\'ee par un moment dipolaire \'electrique par maille 
{\bf P} non nul. Dans un cristal pyro\'electrique, toute op\'eration de sym\'etrie doit pr\'eserver la direction de la polarisation. Le seul axe de rotation possible est donc parall\`ele \`a {\bf P}. En outre il ne peut y avoir de plan miroir perpendiculaire \`a {\bf P}. Ceci n'autorise que les groupes ponctuels $C_n$, $C_{nv}$, et $C_{1h}$, Le groupe $C_1$ est constitu\'e par l'\'el\'ement unit\'e seul, et il est triclinique. Le groupe $C_n$ comporte un axe de rotation d'ordre $n$ ($n$=2, 3 , 4 ,5) et rien d'autre. Le groupe $C_{nv}$ comporte en outre un "miroir" (ou plan de sym\'etrie) parall\`ele \`a cet axe. Le groupe $C_{1h}$ est constitu\'e par l'\'el\'ement unit\'e et un miroir.

Plus simple encore est la bir\'efringence. Elle existe pour tous les cristaux non cubiques. Elle est li\'ee \`a la forme du tenseur de permittivit\'e \'electrique $\epsilon  $ d\'efini par $D=\epsilon E$, o\`u $E$ et $D$ sont respectivement le champ et le d\'eplacement \'electrique. Il y a bir\'efringence si la matrice $\epsilon $ n'a pas ses 3 valeurs propres 
d\'eg\'en\'er\'ees.

Des matrices ? Le physicien en rencontre souvent de bien plus compliqu\'ees, et il a besoin de conna\^{\i}tre leurs valeurs propres et leurs vecteurs propres. Nous allons voir que la th\'eorie des groupes peut l'aider.


\section{Vecteurs propres d'un op\'erateur  et repr\'esentations irr\'eductibles d'un groupe}
\label{sec5}

La th\'eorie des groupes permet de simplifier la recherche et la classification des vecteurs propres (et par cons\'equent des valeurs propres) d'une matrice ou d'un op\'erateur. Il peut s'agir par exemple d'un hamiltonien. 
Un cas bien connu, o\`u l'on fait de la th\'eorie des groupes sans le savoir, est celui d'une matrice $N\times N$ invariante par translation : $M(n,m)=M(n+p,m+p)$ quel que soit l'entier $p$. Nous supposons des conditions aux limites 
p\'eriodiques, c'est \`a dire que $n$ et $m$ sont d\'efinis modulo $N$. On sait que les vecteurs propres ont la forme 
\begin{equation}
u(n)=Const \times \exp(ikn)			
\label{period}
\end{equation}
o\`u $k$ est tels que $\exp(ikN)=1$. C'est une propri\'et\'e famili\`ere, mais \'etonnante : la seule invariance par le groupe des translations d\'etermine les vecteurs propres, quelles que soient les valeurs num\'eriques des \'el\'ements de matrice ! Les valeurs propres s'en d\'eduisent en une ligne, et elles d\'ependent des \'el\'ements de matrice. 
Cette propri\'et\'e est un cas particulier d'une propri\'et\'e g\'en\'erale : si une matrice est $T$ invariante 
par les op\'erations $g$ d'un groupe $G$, tout \'el\'ement $g$ transforme tout vecteur propre $\ket{u}$ en un vecteur propre $T(g) \ket{u}$, o\`u les matrices $T(g)$ forment une repr\'esentation de $G$, 
c'est \`a 
dire que $T(g) T(g')=T(gg')$. Dans l'exemple pr\'ec\'edent, les \'el\'ements de $G$ sont les translations $g_p$ qui transforment $M(n,m)$ en $M(n+p,m+p)$, et $T(g_p)= \exp(ikp)$ est un 
nombre, c'est-\`a-dire  une matrice $1 \times 1$. On dit que les matrices $T(g_p)$ constituent 
une repr\'esentation de dimension 1. Bien s\^ur il y a autant de telles repr\'esentations qu'il y a de nombres $k$, donc il y en a $N$, qu'il faut distinguer par un indice, de sorte que $T_k(g_p)= \exp(ikp)$. 
Avec 2 repr\'esentations $T_k$ et $T_{k'}$ on peut construire une repr\'esentation qui associe \`a $g$ la matrice 
 			
 \begin{equation}
\left[
\begin{array}{cc}
T_k(g) & 0 \\
 0 & T_{k'}(g)
\end{array}
\right]
\label{reduc}
\end{equation}

Une telle repr\'esentation est dite r\'eductible. Dans la recherche des vecteurs propres d'une matrice invariante par un groupe $G$, on peut imposer \`a ces vecteurs de se transformer selon une repr\'esentation irr\'eductible. On trouve ainsi tous les vecteurs propres... \`a condition, \`a la fin du calcul, de combiner ceux qui correspondent \`a chaque valeur propre. Dans l'exemple pr\'ec\'edent, s'il se trouve que la matrice $M(n,m)$ est r\'eelle et sym\'etrique, il faut combiner $u_k (n) = \exp(ikn)$ et            $u_{-k}(n)=\exp(-ikn)$. 

Une repr\'esentation de dimension 1 est \'evidemment irr\'eductible. Les repr\'esentations irr\'educ-tibles 
du groupe des translations s'\'ecrivent donc sans difficult\'es, et pas seulement dans le cas d'un espace unidimensionnel. Dans des cas plus compliqu\'es, la recherche des repr\'esentations irr\'educ-tibles est 
plus difficile, mais simplifie souvent la recherche des vecteurs propres. 

On peut r\'esumer ce paragraphe par les r\`egles fondamentales suivantes : 

1) Pour trouver les vecteurs propres d'un 
op\'erateur invariant par un groupe $G$, on peut se restreindre \`a l'espace des vecteurs qui se transforment selon chaque 
repr\'esentation irr\'eductible. On dit que ces vecteurs propres appartiennent \`a cette repr\'esentation 
irr\'eductible. Cette m\'ethode permet d'obtenir {\it tous} les vecteurs propres.
Cette r\`egle constitue le th\'eor\`eme de d\'eveloppement  (expansion theorem, Streitwolf 1971). Elle ram\`ene 
la diagonalisation d'une grande matrice \`a celle d'un certain nombre de petites matrices, dont chacune correspond
\`a une repr\'esentation irr\'eductible diff\'erente.
La matrice $M$, gr\^ace \`a un choix des
vecteurs de base, prend donc la forme diagonale par blocs: 			
 \begin{equation}
M = \left[
\begin{array}{cccccc}
M_1 & 0 &...& 0 &...& 0 \\
0 & M_2 &...& 0 &...& 0 \\
0 & 0 &...& 0 &...& 0 \\
0 & 0 &...& M_p &...& 0 \\
...& ... &...& ... &...& ... \\
0 & 0 &...& 0 &...& M_r \\
\end{array}
\right]
\label{reduc22}
\end{equation}

2) Pour obtenir chaque petite matrice, la th\'eorie des groupes fournit des recettes. Une telle recette est
bri\`evement d\'ecrite dans l'appendice  \ref{secA1b}.
Pour la justification et pour des  exemples, on peut se reporter \`a Canals (2009) Ballou (2009),  
Rodriguez-Carvajal (2009).

\vskip 1cm
Exercice. Consid\'erer les fonctions d'onde \`a un \'electron dans un potentiel p\'eriodique. Selon le th\'eor\`eme 
de Floquet-Bloch, elles ont la forme $\psi _k (r) = \exp(i {\bf k}.{\bf r}) u_k ({\bf r})$ o\`u 
$u_k ({\bf r})$  a la p\'eriode du potentiel. Relier cette propri\'et\'e aux repr\'esentations 
irr\'eductibles du groupe des translations du r\'eseau. 

\section{Quelques exemples d'applications de la th\'eorie des groupes}
\label{sec6}
\subsection{D\'eg\'en\'erescence d'un hamiltonien \`a spectre discret ou d'une matrice}

Voyons d'abord le cas d'une matrice $2 \times 2$ (le hamiltonien d'un spin 1/2 par exemple) dont les 
\'el\'ements sont $a_{ij}$. Pour qu'elle soit d\'eg\'en\'er\'ee, il faut qu'elle soit un multiple de la matrice unit\'e. Il faut donc remplir les conditions $a_{11}=a_{22}$ et $a_{12}=a_{21}=0$. M\^eme si la matrice est hermitique (cas d'un hamiltonien) ces conditions sont difficiles \`a remplir et exigent de faire varier plus d'un param\`etre.... A moins que les deux vecteurs propres n'appartiennent \`a des repr\'esentations diff\'erentes. Dans ce cas, $a_{12}$ et $a_{21}$ sont automatiquement nuls et la 
d\'eg\'en\'erescence appara\^{\i}t quand $a_{11}=a_{22}$.

\begin{figure}[t]
\centering
 \includegraphics*[width=70mm, ]{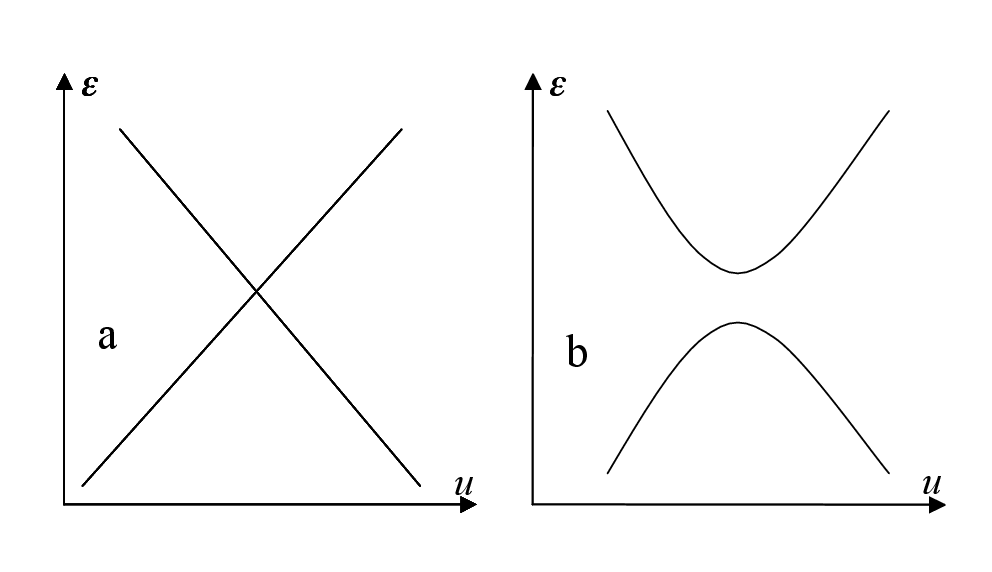}
\caption{a) Si deux valeurs propres d'une matrice appartiennent \`a 2 repr\'esentations diff\'erentes, la variation 
d'un param\`etre u permet de les rendre \'egales.  b) Si les  deux valeurs propres appartiennent \`a la m\^eme 
repr\'esentation, elles ne sont en g\'en\'eral jamais \'egales.}
\label{fig2}
\end{figure}

Le cas d'une matrice plus grande est plus compliqu\'e, mais analogue : la variation d'un seul 
param\`etre ne fait en 
g\'en\'eral pas appara\^{\i}tre de d\'eg\'en\'erescence exacte (Fig. \ref{fig2}) sauf pour des valeurs propres appartenant \`a des repr\'esentations diff\'erentes. 

\subsection{Effet Jahn-Teller}

Supposons que les valeurs propres dont il vient d'\^etre question soient les \'energies possibles d'un \'electron 
localis\'e dans un cristal, dont la sym\'etrie est assez \'elev\'ee pour que les 2 valeurs propres 
soient d\'eg\'en\'er\'ees. 
D'apr\`es ce qui vient d'\^etre dit, cette d\'eg\'en\'erescence ne peut gu\`ere se produire que si les vecteurs propres appartiennent \`a 2 repr\'esentations diff\'erentes. Si le cristal subit une distorsion,  la d\'eg\'en\'erescence 
dispara\^{\i}t avec la sym\'etrie. Cela entra\^{\i}ne une augmentation d'une des valeurs propres et une diminution de l'autre. 
L'\'electron choisit la plus basse valeur propre, de sorte que la distorsion est \'energ\'etiquement favorable (Amara 2009). 
Il faut \'evidemment qu'il n'y ait qu'un \'electron pour les deux \'etats. 

\subsection{Spectre de phonons d'un cristal}

Les fr\'equences des phonons de vecteur d'onde {\bf q}  sont les valeurs propres d'une matrice finie dont les \'el\'ements ne sont en g\'en\'eral connus qu'avec une pr\'ecision m\'ediocre. 
Mais la sym\'etrie de la matrice est connue. Si le vecteur {\bf q} 
est quelconque, la matrice n'a aucune sym\'etrie et la th\'eorie des groupes ne sert \`a rien. Par contre, elle peut simplifier consid\'erablement le calcul dans le cas o\`u {\bf q}  est, 
par exemple, sur un axe de sym\'etrie. En particulier, 
on peut alors savoir quelles sont les d\'eg\'en\'erescences (Kreisel 2009). 

\subsection{Structures magn\'etiques}

Quand les atomes d'un cristal portent un moment magn\'etique, ces moments s'ordonnent g\'en\'eralement \`a basse 
temp\'erature. 
La structure magn\'etique qui en r\'esulte se superpose \`a la structure cristallographique. Deux cas sont possibles,  
dont des exemples sont d\'ecrits par  Bour\'ee (2009).

Premier cas: la structure magn\'etique et  la structure cristallographique ont une periodicit\'e commune (avec une maille qui peut \^etre plus grande que la maille cristallographique paramagn\'etique de haute  temp\'erature). 
Les op\'erations de sym\'etrie qui conservent la structure magn\'etique et  la structure cristallographique forment un groupe. Les groupes  magn\'etiques ainsi obtenus peuvent \^etre  catalogu\'es \`a la fa\c con   des 
groupes d'espace habituels (Rodrigez-Carvajal 2009). Leur nombre est fini.

Deuxi\`eme cas: la structure magn\'etique a une p\'eriode incommensurable avec le r\'eseau cristallin. La situation la plus favorable est le voisinage d'une  transition continue du paramagn\'etisme \`a une phase magn\'e tiquement ordonn\'ee. D'apr\`es la th\'eorie de Landau des transitions de phase (Toledano 2009) la structure magn\'etique doit correspondre \`a un vecteur propre d'une certaine matrice, la matrice susceptibilit\'e paramagn\'etique g\'en\'eralis\'ee (Schweizer et al. 2007), qui d\'efinit la r\'eponse lin\'eaire des moments \`a un champ magn\'etique modul\'e dans l'espace. La th\'eorie des groupes permet, comme on l'a expliqu\'e au paragraphe \ref{sec5}, de faciliter cette recherche. 
Une m\'ethode bas\'ee sur les repr\'esentations a \'et\'e propos\'ee initialement par Bertaut (1971) 
puis perfectionn\'ee par Schweizer (2005, 2006, 2007, 2009). A basse temp\'erature, diverses complications peuvent se produire. D'abord, des harmoniques apparaissent \`a cause de l'anisotropie magn\'etocristalline. En outre, un verrouillage (lock-in) des deux p\'eriodes (magn\'etique et cristallographique) est fr\'equent. La structure magn\'etique  devient alors commensurable avec le r\'eseau cristallin,  avec une p\'eriodicit\'e qui peut \^etre tr\`es longue. Quand on fait varier un param\`etre (pression, temp\'erature...) on peut avoir, en principe, une succession  de verrouillages prenant l'aspect d'un ``escalier du diable'' (Axel et Aubry 1981) ou d'un escalier ``inoffensif'' (Villain et Gordon 1980).


\subsection{L'h\'elium 3 superfluide et la supraconductivit\'e inhabituelle}

La supraconductivit\'e est caract\'eris\'ee par une valeur non nulle de 
$\Delta_{\sigma \sigma '} ({\bf k}) =\langle c_{k \sigma } c_{-k \sigma' } \rangle$ 
o\`u l'op\'erateur $c_{k \sigma }$ annihile un \'electron d'impulsion $\hbar {\bf k}$
et de spin $\sigma = \pm 1/2$. 
   
Dans la th\'eorie  classique de la supraconductivit\'e \'elabor\'ee par  Bardeen, Cooper et Schrieffer, 
les paires de Cooper sont dans un \'etat de spin singulet et le  param\`etre d'ordre
$\Delta_{\sigma \sigma '} ({\bf k}) $ est ind\'ependant la direction  ${\bf k}/k$.
Il est pratiquement  nul loin de la surface de Fermi, o\`u il a une valeur  $\Delta $.
Ceci est une bonne description de la plupart des supraconducteurs usuels comme Nb. 

Dans les supraconducteurs inhabituels ("non-conventional"), le  param\`etre d'ordre
$\Delta_{\sigma \sigma '} ({\bf k}) $ peut d\'ependre de ${\hat k}={\bf k}/ k$ (Houzet 2009).
De plus, les paires de Cooper peuvent \^etre dans l'\'etat triplet de spin. 
Quand les op\'erations du groupe ponctuel $\mathcal{G}_0$ du cristal agissent sur {\bf k},
la fonction $\Delta_{\sigma \sigma '} ({\bf k}) $
doit se transformer comme une repr\'esentation irr\'eductible  de $\mathcal{G}_0$.
Ainsi, la th\'eorie des groupes fournit une classification des divers types de supraconductivit\'e possibles.
Cette classification facilite l'\'etude exp\'erimentale.

Dans  $^3$He superfluide, de fa\c con analogue, il se forme des paires de  fermions de vecteurs d'onde  {\bf k} et 
{-\bf k}. 
Les paires sont dans l'\'etat triplet.
La fonction d'onde orbitale doit se transformer comme une   repr\'esentation irr\'eductible   
$Y_{\ell m}(\theta , \varphi )$ du groupe des
rotations, o\`u $\theta  $ et $\varphi $ sont les angles d'Euler de  ${\hat k}$. 
La condition usuelle sur l'antisym\'etrie de la fonction d'onde des fermions impose que $\ell$
soit impaire. En pratique, $\ell=1$.

\vskip 5mm

L'\'enum\'eration qui pr\'ec\`ede est succincte avec exc\`es.
Elle sera compl\'et\'ee et explicit\'ee par les expos\'es cit\'es en r\'er\'erence
et par  des manuels tels que ceux de Meijer \& Bauer (2004), de Hamermesh (1962)  ou de Streitwolf (1971) 
dont le volume modeste et le formalisme simple rendent la lecture agr\'eable.

\section{Fonctions d'onde et groupe des rotations}
\label{sec7}

L'espace est invariant par translation et par rotation. Nous avons vu que le groupe des translations est trop simple pour que la th\'eorie des groupes soit indispensable. Voyons  donc quelles informations peut apporter l'invariance par rotation. Le probl\`eme se pose par exemple si on veut calculer la fonction d'onde d'un \'electron soumis \`a l'attraction d'un noyau situ\'e en un point fixe O ; c'est le cas de l'atome d'hydrog\`ene. Ce probl\`eme est bien connu, mais il est utile de le r\'eexaminer en insistant sur l'invariance par rotation. D'apr\`es la r\`egle fondamentale, nous pouvons simplifier la recherche des fonctions d'onde (solutions de l'\'equation de Schr\"odinger) en cherchant les repr\'esentations irr\'eductibles du groupe des rotations. C'est, en r\'ealit\'e, ce qui est fait dans les manuels de m\'ecanique quantique, mais souvent sans le dire. 

Les repr\'esentations irr\'eductibles de dimension $(2\ell+1)$ impaire du groupe des rotations peuvent s'obtenir en prenant pour vecteurs de base les fonctions 
\begin{equation}
Y_{lm}(\theta, \varphi ) =  B_{lm} P_{lm}(\cos \theta  )\exp(im \varphi )
\label{harmo1}
\end{equation}
o\`u les nombres $B_{lm}$ sont des constantes de normalisation et
\begin{equation}
 P_{lm}(w)=(1-w^2)^{|m|/2} d ^{|m|}P_l(w)/dw^{|m|}
\label{harmo2}
\end{equation}
o\`u $P_l(w)$ est le polyn\^ome de Legendre d'ordre $l$
\begin{equation}
 P_l(w)=(2^{-l}/l !)d^l(w^2-1)^l/dw^l	
\label{harmo3}
\end{equation}
	
La th\'eorie des groupes nous dit que les solutions de l'\'equation de Schr\"odinger dans un potentiel invariant par rotation ont la forme
\begin{equation}
\psi _{lm}(r, \theta, \varphi ) = R_l(r) Y_{lm}(\theta, \varphi )
\label{harmo5}
\end{equation}
et qu'il n'y a plus qu'\`a d\'eterminer la fonction radiale $R_l(r)$.  Cette factorisation est, bien s\^ur,  dans les manuels, mais elle est souvent pr\'esent\'ee comme un miracle alors que la th\'eorie des groupes permet de la pr\'evoir.
D'autre part, les manuels introduisent g\'en\'eralement les harmoniques sph\'eriques $Y_{lm}(\theta, \varphi )$ comme instruments d'\'etude du moment angulaire et ne font pas toujours la liaison avec les rotations. 
La fin de ce paragraphe est consacr\'ee \`a cette relation.
Il s'agit de savoir comment une 
fonction $f(x,y,z) $ (une fonction d'onde par exemple) se transforme par une rotation des axes, 
par exemple autour de l'axe $z$ qui est donc
 invariant. Si $\varphi $ est l'angle de  rotation, les nouvelles coordonn\'ees sont
$x'= x\cos \varphi+y\sin \varphi$ et $ y'=-x\sin  \varphi+y\cos \varphi$. La fonction $f(x,y,z) $ se
transforme en une fonction de $x'$, $y'$ et $z$ qui, pour  $\varphi $ petit, peut s'\'ecrire 
$$
f(x'\cos \varphi-y'\sin \varphi, x'\sin  \varphi+y'\cos \varphi, z')
=f(x'-y' \varphi, x' \varphi+y', z') = (1+  \rho_z) f(x', y', z') 
$$
o\`u l'op\'erateur de rotation  infinit\'esimal  $\rho_z$ est $\rho _z = x \partial y - y \partial x$.
Il co\"{i}ncide avec   $i L_z/\hbar$, o\`u ${\bf L}$ est l'expression du moment angulaire orbital en m\'ecanique quantique. 
Des operateurs de rotation  infinit\'esimaux  $\rho_x$ et $\rho_y$  
autour de $x$ et $y$ peuvent \^etre obtenus de fa\c con analogue. 
L'identit\'e  $\rho _\alpha  = i L_\alpha /\hbar$ est valable pour les  3 composantes $\alpha =x,y,z$.
On en d\'eduit les r\`egles de   commutation 
 
 \begin{equation}
[\rho _x,\rho _y]= -\rho _z\;\;\;\;\;;\;\;\;\;[\rho _y,\rho _z]= 
-\rho _x\;\;\;\;\;;\;\;\;\;[\rho _z,\rho _x]= -\rho _y  
\label{rota1}
\end{equation}

On verra que la relation entre op\'erateur de rotation  et moment angulaire  
est un cas particulier du th\'eor\`eme de  Noether \'enonc\'e au paragraphe \ref{sec12}. 

Dans le pr\'esent paragraphe, on a introduit \`a partir de la th\'eorie des groupes, et somme toute de 
la g\'eom\'etrie, des concepts et des r\'esultats (harmoniques sph\'eriques, 
r\`egles de commutation) qui sont g\'en\'eralement d\'eduits de la m\'ecanique quantique.
Or ce raisonnement g\'eom\'etrique est plus simple et plus \'el\'ementaire, puisque
l'on n'a pas fait usage de  la relation de De Broglie 
${\bf p}=i\hbar \nabla$, hypoth\`ese physique \`a laquelle nous sommes habitu\'es, mais tout de m\^eme moins \'el\'ementaire que 
la simple g\'eom\'etrie.
L'approche g\'eom\'etrique est aussi plus g\'en\'erale et plus \'el\'egante.

Les r\`egles de commutation  (\ref{rota1}) sont des propri\'et\'es essentielles des op\'erateurs de rotation. 
Elles sont le  point de d\'epart in\'evitable de la recherche des  repr\'esentations de SO3. Cette
recherche m\`ene, comme on va le voir, au concept de spineur.

\section{Les repr\'esentations irr\'eductibles de dimension paire du groupe des rotations et le spin.}
\label{sec9}

La formule (\ref{harmo1}) ne fournit  que des repr\'esentations irr\'eductibles   de dimension impaire du groupe des rotation SO3. 
Le pr\'esent paragraphe est une qu\^ete des repr\'esentations irr\'eductibles   de dimension paire de SO3. 
En fait on se limitera \`a chercher une repr\'esentations irr\'eductible   de dimension 2, ce qui est suffisant
pour les besoins des sciences physiques.

Une telle repr\'esentation doit contenir trois  matrices 2$\times$2 associ\'ees aux rotations infinit\'esimales 
autour des axes $x$, $y$, $z$. Ces matrices doivent v\'erifier des relations de commutation analogues \`a (\ref{rota1}). 
Comme il r\'eulte des manuels de m\'ecanique quantique, de telles matrices sont les matrices de Pauli

 \begin{equation}
\sigma _x =\left[
\begin{array}{cc}
0 & 1 \\
1 & 0
\end{array}
\right]
\;\;\;\;\;\;;\;\;\;\;\;\;
\sigma _y =\left[
\begin{array}{cc}
0 & -i \\
i & 0
\end{array}
\right]
\;\;\;\;\;\;;\;\;\;\;\;\;
\sigma _z =\left[
\begin{array}{cc}
1 & 0 \\
0 & -1
\end{array}
\right]
\label{harmo6}
\end{equation}
qu'il convient seulement de multiplier par  $i/2$. On peut ainsi esp\'erer que les 
matrices\footnote{L'\'egalit\'e (\ref{rota2}) peut \^etre v\'erifi\'ee en d\'eveloppant 
l'exponentielle en s\'erie et  en utilisant les relations d'anticommutation des matrices  de Pauli.}

\begin{equation}
\exp[(i \varphi  /2)(\alpha \sigma _x  + \beta \sigma _y + \gamma  \sigma _z) ]
=1 \cos \varphi /2 + i (\alpha \sigma _x  + \beta \sigma _y + \gamma  \sigma _z) \sin \varphi /2 )
\label{rota2}
\end{equation}
forment une repr\'esentation du groupe des rotations SO3.

Dans le langage des physiciens, il en est bien ainsi. La matrice (\ref{rota2}) correspond \`a une rotation d'angle $\varphi   $
autour d'un axe parall\`ele au vecteur unit\'e de composantes $\alpha, \beta, \gamma$. 

Malheureusement, la formule (\ref{rota2}) a une \'etrange propri\'et\'e. La  rotation d'angle 
$2 n \pi $  autour de n'importe quel axe est \'evidemment l'identit\'e.
Or, si $\varphi = 2 n \pi $,  la matrice (\ref{rota2})  est \'egale \`a 1 si $n$ est pair (ce qui est fort bien) mais \'egale \`a -1 si $n$ est impair. Plus g\'en\'eralement, la relation (\ref{rota2}) associe  \`a toute rotation  deux matrices 
$A$ and $-A$. Ceci n'est pas autoris\'e par la d\'efinition d'une repr\'esentation accept\'ee par les math\'ematiciens.

Nous dirons que les matrices (\ref{rota2}) forment une ``repr\'esentation spinorielle'' de SO3. 
Une  repr\'esentation spinorielle  est bivalu\'ee, et par suite, {\it stricto sensu}, ce n'est pas une repr\'esentation.
Il nous arrivera cependant, par la suite, de l'appeler ainsi pour simplifier.

Une  repr\'esentation spinorielle  agit sur des matrices colonnes \`a deux \'el\'ements qu'on appelle des  spineurs. 
Le terme `spineur' se justifie parce que ces objets sont utilis\'es pour la  description math\'ematique d'un spin 1/2. Cependant, le concept de spineur avait \'et\'e invent\'e par le math\'ematicien \'Elie Cartan d\`es 1913, 9 ans avant 
la d\'ecouverte exp\'erimentale du spin de  l'electron par Stern and Gerlach en 1922.
Les math\'ematiques \'etaient d\'ej\`a pr\^etes \`a \^etre utilis\'ees par les  physiciens.

Au paragraphe \ref{sec7}, on avait utilis\'e la th\'eorie des groupes pour obtenir  
les vecteurs propres d'un op\'erateur donn\'e agissant sur un espace vectoriel connu. Dans ce paragraphe,
la th\'eorie des groupes a fonctionn\'e autrement. 
La nature spinorielle de la fonction d'onde n'\'etait   {\it a priori} pas connue, 
c'est la th\'eorie des groupes qui a permis de la d\'ecouvrir. Certes, beaucoup de manuels de 
m\'ecanique quantique n'y font pas appel, mais sans
la th\'eorie des groupes l'apparition des spineurs devient artificielle.

\section{Fonctions d'onde relativistes d'un \'electron.}
\label{sec10}

La relativit\'e fait intervenir, comme on l'a vu, un espace \`a 4 dimensions, dans lequel les changements d'axes  combinent  
le temps $t$ (ou plut\^ot $x^0=ct$) et les 3 coordonn\'ees d'espace $x^\alpha $ ($\alpha =1,2,3$).

Alors que la recherche de la fonction d'onde d'une particule non relativiste nous avait amen\'es \`a 
rechercher les repr\'esentations irr\'eductibles du groupe des rotations, c'est maintenant les 
repr\'esentations du groupe de Lorentz qui nous int\'eressent. Nous nous bornerons ici au cas 
d'une particule de spin 1/2 comme l'\'electron.  La repr\'esentation appropri\'ee est \`a 2 dimensions, et spinorielle.
Elles fait appel aux matrices de Pauli et ne sera pas donn\'ee ici. Par contre, nous insisterons sur une
difficult\'e qui n'apparaissait pas en physique quantique non-relativiste.

Il se trouve qu'il n'y a pas moyen de d\'efinir une fonction d'onde qui se transforme selon la repr\'esentation spinorielle irr\'eductible de dimension 2 du groupe de Lorentz. Comme on le montre dans l'appendice
\ref{Cartan}, il faut introduire une  repr\'esentation spinorielle de dimension 4,
 qui est r\'eductible et s'obtient en combinant la repr\'esentation  irr\'eductible de dimension 2 avec une 
repr\'esentation analogue, mais non \'equivalente
 (Cartan 1938).   On obtient ainsi une fonction d'onde spinorielle \`a 4 composantes qui 
v\'erifie l'\'equation
 de Dirac:

 $$
\hbar  \gamma^\mu \partial_\mu  \psi + mc \psi =0
$$
o\`u il faut sommer sur $\mu $ (convention d'Einstein). Les matrices 4$\times$4   $\gamma ^\mu  $ 
v\'erifient les r\`egles   d'anticommutation 
$\{\gamma ^\mu ,\gamma ^\nu \}=2 g^{\mu  \nu }$, o\`u 
 \begin{equation}
g^{\mu  \nu } =g^\mu \delta ^{\mu \nu }
\label{gg}
 \end{equation}
 et\footnote {La quantit\'e $g^{\mu  \nu }$ est un tenseur, qui garde la m\^eme forme diagonale dans tous les syst\`emes 
 d'axes reli\'es par une
transformation de Lorentz. Au contraire, $\delta ^{\mu  \nu }$ n'est pas un tenseur et  $g^\mu$,
d\'efini par (\ref{ggg}), n'est pas  vecteur. La relation (\ref{ggg}) peut \^etre vraie pour un choix particulier des axes, 
 mais elle n'est pas pr\'eserv\'ee par une transformation
de Lorentz. On dit que la quantit\'e $g^\mu$ n'est pas covariante. Pour cette raison, on ne l'introduit en g\'en\'eral pas. }  
 \begin{equation}
g^{1}= g^{2}= g^{3}=1 \;\;\;; \;\;\;g^{0}=-1
\label{ggg}
\end{equation}

Un choix possible est 
 \begin{equation}
\gamma  ^\alpha  =\left[
\begin{array}{cc}
0 & - i \sigma _\alpha  \\
i \sigma _\alpha   & 0
\end{array}
\right]
\;\;\;\;\;\;(\alpha =1,2,3);\;\;\;\;\;\;
\gamma  ^0 =i \left[
\begin{array}{cc}
1 & 0 \\
0  & -1
\end{array}
\right]
\label{harmoA2}
\end{equation}
o\`u les blocs sont des matrices 2$\times$2. Les $\sigma _i $ sont les matrices de Pauli.

\section{L'\'electromagn\'etisme et l'\'electrodynamique: sym\'etrie de jauge}
\label{sec11}

Si une particule mat\'erielle a une fonction d'onde qu'on peut souvent supposer scalaire, il n'en est pas ainsi du photon. La lumi\`ere est caract\'eris\'ee par un champ \'electrique et un champ magn\'etique, qui sont des vecteurs. Dans ce cas, le groupe des rotations \`a 3 dimensions et ses repr\'esentations ne sont gu\`ere utiles. Le groupe des rotations est important dans le cas d'un \'electron qui tourne autour d'un noyau immobile. Les photons n'ont pas l'habitude de tourner autour d'un centre. 

Bien entendu, l'\'electromagn\'etisme n\'ecessite l'emploi de la relativit\'e. Fort heureusement, la relativit\'e restreinte suffit. Dans ce cadre, le champ \'electromagn\'etique est caract\'eris\'e par un tenseur antisym\'etrique $ F_{\mu \nu } $  \`a $4\times 4$ composantes, dont 6 sont non nulles et ind\'ependantes (champ \'electrique 
et champ magn\'etique). Mais ce tenseur peut s'exprimer en fonction du potentiel vecteur $A_\mu $ :
\begin{equation}
F_{\mu \nu } = \partial _\mu A_\nu - \partial _\nu A_\mu 
\label{electro1}
\end{equation}

Comme il vaut mieux avoir une quantit\'e \`a 4 composantes qu'une quantit\'e \`a 6 composantes, l'emploi de $A$ \`a la place de $F$ est in\'evitable (au moins en m\'ecanique quantique).  Mais cela implique une difficult\'e: En effet, la quantit\'e physique mesurable n'est pas le potentiel vecteur $A$, mais  le champ \'electromagn\'etique, et celui-ci n'est pas modifi\'e si on remplace $A $ par 
\begin{equation}
A'_\mu =A_\mu - g_\mu \partial _\mu \Lambda ({\bf r},t)
\label{electro2}
\end{equation}
o\`u   $g_\mu $ est d\'efini (\ref{ggg})\footnote{Comme on l'a remarqu\'e pr\'ec\'edemment, $g^\mu$ n'est pas covariant. Par suite, $\Lambda ({\bf r},t)$ n'est pas forc\'ement  convariant non plus. La `jauge de Coulomb' qu'on utilise beaucoup, ne l'est pas.}.
Le choix arbitraire de $\Lambda $ s'appelle une jauge. Comme la quantification du champ \'electromagn\'etique (7) ou (8) conduit \`a des photons, ceux-ci sont appel\'es bosons de jauge.
Si l'on a une particule charg\'ee (\'electron) qui interagit avec le champ \'electromagn\'etique, le choix de la jauge reste arbitraire, mais un changement de la jauge implique \'egalement un changement de la phase de la fonction d'onde (Voir appendice \ref{secA2}). Celle-ci est donc multipli\'ee par un nombre $\exp(i\varphi)$. Ce facteur engendre   un groupe, le groupe U1. 

Il est \`a remarquer que dans un superfluide ou un supraconducteur, il appara\^{i}t aussi une {\it phase} qui est li\'ee aux op\'erateurs de champ. Le hamiltonien est invariant par un changement global de cette phase. Cette invariance est bris\'ee dans un superfluide ou un supraconducteur. On d\'ecrit souvent ce ph\'enom\`ene comme une brisure de l'invariance de jauge. Il y a en effet une analogie avec la jauge en \'electrodynamique, puisque c'est aussi le groupe U1 qui intervient, mais le ph\'enom\`ene physique est diff\'erent et les photons ne jouent aucun r\^ole dans un superfluide.

Les transformations de jauge de l'\'electrodynamique forment un groupe. Cette remarque n'avance cependant pas \`a grand' chose car le groupe ab\'elien U(1) est fort simple et ne n\'ecessite pas la machinerie lourde de la th\'eorie des groupes. D'excellents manuels d'\'electrodynamique quantique n'\'evoquent m\^eme pas le concept de groupe de jauge. Cependant, l'\'electrodynamique fournit une introduction didactique \`a des th\'eories de jauge bien plus compliqu\'ees, n\'ecessaire pour d\'ecrire les particules \'el\'ementaires autres que l'\'electron. Avant d'en parler, il convient d'\'enoncer 
2 th\'eor\`emes g\'en\'eraux, qui servent en  physique de la mati\`ere condens\'ee comme en physique des particules.

\section{Deux th\'eor\`emes g\'en\'eraux}
\label{sec12}
\subsection{Th\'eor\`eme de  Noether (1915) }
\label{ssec12a}

A tout groupe continu de sym\'etries peut \^etre associ\'ee une quantit\'e conserv\'ee, et vice-versa (Noether 1918, Byers 1996). Ainsi la conservation de la quantit\'e de mouvement est la cons\'equence de la sym\'etrie de translation dans l'espace. De m\^eme, la conservation de l'\'energie cin\'etique est li\'ee \`a l'invariance par translation dans le temps. La conservation du moment cin\'etique provient de l'invariance par rotation.

\subsection{Th\'eor\`eme de Goldstone}
\label{ssec12b}
Le th\'eor\`eme de Goldstone affirme l'apparition d'excitations de "gap" nul quand une sym\'etrie continue est bris\'ee. Consid\'erons par exemple un cristal. Il brise la sym\'etrie continue de translation de l'espace. Les excitations correspondantes sont les phonons acoustiques, dont la fr\'equence $\omega _q$ est li\'ee au vecteur d'onde {\bf q}, pour $q$ petit, par la relation lin\'eaire $\omega _q= \alpha q$. Leur \'energie s'annule donc quand $q $ tend vers 0. Le th\'eor\`eme de Goldstone (Goldstone1961, Goldstone et al. 1962) affirme que cette propri\'et\'e est g\'en\'erale : toute brisure de sym\'etrie continue s'accompagne de l'apparition d'excitations (g\'en\'eralement des bosons) dont l'\'energie peut tendre vers 0. 
Dans le cas du magn\'etisme, la brisure de sym\'etrie de rotation des spins entra\^{\i}ne 
l'apparition de magnons \`a basse temp\'erature. Mais attention ! La sym\'etrie de rotation des spins 
est bien pr\'esente (\`a haute temp\'erature) dans certains mod\`eles comme le mod\`ele de Heisenberg, 
mais ces mod\`eles qui n\'egligent l'anisotropie magn\'etocristalline ne sont qu'approch\'es, et l'\'energie des magnons ne s'annule pas strictement.

Quant \`a la supraconductivit\'e, elle est trait\'ee dans l'appendice \ref{secA4}.

Si les physiciens de la mati\`ere condens\'ee se sont appropri\'es le th\'eor\`eme de Goldstone, il est bon de rappeler que l'id\'ee de Goldstone \'etait de l'appliquer aux particules \'el\'ementaires. 
Sa d\'emonstration \'etait donc relativiste... et utilisait un bagage technique appr\'eciablement plus perfectionn\'e que celui dont ont besoin les physiciens de la mati\`ere condens\'ee. Or, en relativit\'e, l'\'energie d'une particule est $E=c(m^2 c^2+p^2) ^{1/2} $; si la masse au repos $m$ est finie, cette \'energie s'\'ecrit, pour $p$ petit, $E=m c^2+p^2/(2m) $; mais si la masse est nulle, alors $E=cp=c\hbar q$. Les physiciens des particules \'el\'ementaires aiment donc formuler le th\'eor\`eme de Goldstone en disant que la brisure d'une sym\'etrie continue entra\^{\i}ne l'existence de bosons de masse nulle\footnote{En physique de la mati\`ere condens\'ee aussi, les quasi-particules sans masse comme les phonons ont un spectre lin\'eaire, $\omega _q=\alpha q$, alors que des particules de masse non nulle, par exemple les \'electrons, ont un spectre quadratique.} 

Masse nulle ou non nulle ? C'est ce qui d\'etermine la port\'ee de l'interaction transport\'ee par des 
"bosons de jauge"). Consid\'erons d'abord le champ \'electromagn\'etique, v\'ehicul\'e par des photons, 
dont la masse est nulle. Une charge \`a l'origine O produit \`a distance $r$ un potentiel \'electrique $V(r)$ proportionnel \`a $1/r$, solution (pour $r$ non nul) de l'\'equation de Laplace $\nabla ^2V=0$. Mais si le potentiel est d\^u \`a des particules de masse $m$ non nulle, il doit v\'erifier l'\'equation de Schr\"odinger $\nabla ^2V=\kappa ^2V$ avec $\kappa =mc/\hbar$(Yukawa 1949). La solution est $V(r)=Const \times \exp(-\kappa r)/r$ , donc \`a courte port\'ee.

En fait, le th\'eor\`eme de Goldstone fut entre 1960 et 1970 pour les physiciens des particules \'el\'ementaires une source de perplexit\'e comme le verrons.

\section{Les particules \'el\'ementaires}
\label{sec13}

\subsection{Les sym\'etries qui sembleraient s'imposer : parit\'e, etc.}
\label{ssec13}

En physique de la mati\`ere condens\'ee, le nombre d'\'electrons est suppos\'e constant, tout comme le nombre de noyaux de chaque \'el\'ement. La seule interaction consid\'er\'ee est l'interaction \'electromagn\'etique (et quelquefois la pesanteur). C'est une bonne approximation sur la terre, m\^eme si quelquefois la pr\'esence d'une substance radioactive nous oblige \`a faire un peu de physique nucl\'eaire. Cette science met en jeu une interaction d'une autre nature, l'interaction nucl\'eaire forte, qui lie les nucl\'eons dans le noyau. L'ordre de grandeur de l'\'energie \`a consid\'erer est plus \'elev\'e. Par exemple l'\'energie des fragments de fission de l'uranium se mesure en MeV alors que les niveaux \'electroniques dans un atome se mesurent en eV ou en KeV. 

La physique des particules \'el\'ementaires s'int\'eresse \`a des \'energies plus \'elev\'ees encore, qui se mesurent en GeV. Les r\`egles de sym\'etrie auxquelles nous sommes habitu\'es r\'esistent-elles \`a ces \'energies colossales ? 
Ainsi les \'equations du mouvement que nous \'ecrivons \`a basse \'energie sont invariantes par le renversement du temps $T$ et par le renversement $P$ des coordonn\'ees d'espace. En est-il de m\^eme pour la physique beaucoup plus riche dans laquelle on introduit les interactions nucl\'eaires en plus de l'interaction \'electromagn\'etique ?

La r\'eponse est oui, dans la mesure o\`u on ne consid\`ere que les interactions fortes, et \'electromagn\'etiques. En outre, il y a invariance par rapport \`a un autre op\'erateur $C$ inconnu en physique de la mati\`ere condens\'ee : c'est l'op\'erateur qui change une particule en antiparticule (par exemple un \'electron en positon). Cette invariance limite les r\'eactions possibles et simplifie beaucoup la th\'eorie de l'interaction forte et ses applications. 

Mais, l'interaction forte ne d\'ecrit pas tous les ph\'enom\`enes, et certains (comme la d\'esint\'egration $\beta $) font intervenir une autre interaction, dite faible. Et cette derni\`ere n'est pas invariante par $P$, 
ni par $C$, ni par $T$. On crut quelque temps qu'elle est invariante 
par le produit $CP$ (qui transforme par exemple un proton de spin 1/2 en un antiproton de spin -1/2) mais ce n'est pas tout \`a fait vrai non plus. Il reste le produit $CPT$. L'invariance par $CPT$ est garantie par un th\'eor\`eme d\'emontr\'e en 1954 par L\"uders et Pauli \`a partir d'hypoth\`eses quasiment inattaquables.

\subsection{Des id\'ees plus surprenantes: isospin, etc.}
\label{ssec13'}
Les th\'eoriciens des hautes \'energies ont introduit bien d'autres sym\'etries, qui leur paraissent indispensables pour simplifier le foisonnement de particules. Mais faut-il parler davantage de particules \'el\'ementaires dans une \'ecole consacr\'ee \`a la sym\'etrie dans la mati\`ere condens\'ee ?
Une premi\`ere raison de le faire est qu'il y a eu, \`a certaines \'epoques, une forte et fructueuse interaction entre les physiciens des deux bords, et que certains comme Goldstone, Wilson, Anderson, Parisi, Yang, Lee, se sont illustr\'es dans les deux branches. 

Une deuxi\`eme raison est que, dans une \'ecole sur la sym\'etrie tenue en 2009, il est difficile d'ignorer le prix Nobel de physique 2008, d\'ecern\'e aux Japonais Yoichiro Nambu, Makoto Kobayashi et Toshihide Maskawa pour leurs travaux sur la sym\'etrie dans la physique des particules \'el\'ementaires.
Les sym\'etries qu'utilisent les th\'eoriciens des particules \'el\'ementaires ne s'imposent pas de fa\c con imm\'ediate comme celles qui apparaissent en physique de la mati\`ere condens\'ee. Le pionnier en la mati\`ere fut Heisenberg, qui vers 1932 sugg\'era qu'une sym\'etrie existe entre le proton et le neutron.... \`a condition d'ignorer l'interaction \'electromagn\'etique, qui est faible \`a l'\'echelle du noyau. En ne tenant compte que de l'interaction nucl\'eaire forte, le proton et le neutron peuvent en effet \^etre consid\'er\'es comme deux \'etats d'une m\^eme particule, le nucl\'eon. De m\^eme que les deux \'etats d'un ferromagn\'etique sont distingu\'es par les spins des \'electrons, on d\'efinit une variable analogue au spin, qu'on appelle l'isospin. Comme le spin de l'\'electron, il a 2 valeurs propres possibles. Le lagrangien de l'interaction forte, telle que Heisenberg la concevait, devait \^etre invariant par SU(2), ce qui m\`ene bien \`a 2 valeurs propres pour la repr\'esentation fondamentale.

\subsection{Quarks and gluons}
\label{ssec13''}
\begin{figure}[t]
\centering
 \includegraphics*[width=70mm, ]{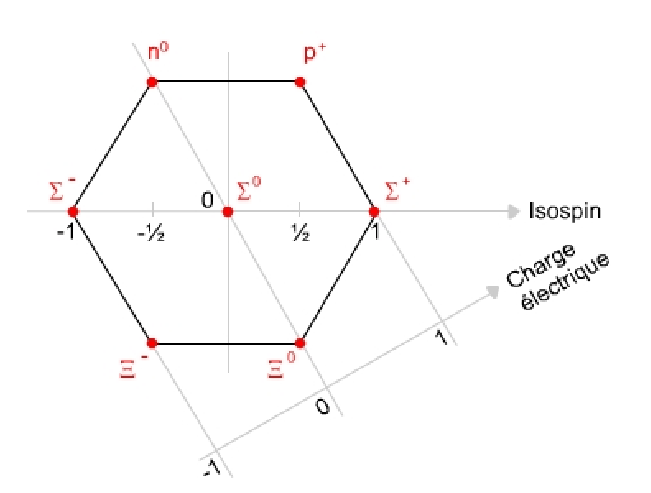}
\caption{Une image de la sym\'etrie entre particules. Si l'isospin, imagin\'e par Heisenberg, introduisait une sym\'etrie entre le proton et le neutron, on peut compl\'eter (ou compliquer ?) le sch\'ema par l'introduction de particules 
$\Sigma $ et $\Xi $, qui sont elles aussi des combinaisons de quarks.}
\label{fig3}
\end{figure}

H\'elas ! La th\'eorie des particules \'el\'ementaires \'evolua. Les protons et neutrons perdirent leur dignit\'e de particules \'el\'ementaires et se d\'ecompos\`erent en 3 quarks. La premi\`ere th\'eorie des quarks fut \'elabor\'ee notamment par Gell-Mann dans le troisi\`eme quart du vingti\`eme si\`ecle. Elle est d\'ecrite 
avec un tout petit peu plus de d\'etails dans l'appendice \ref{secA3}. Les quarks interagissent par \'echange de gluons comme les \'electrons interagissent par \'echange de photons. Les gluons, comme les photons, ont une masse nulle ; comme eux, ils ont une invariance locale vis \`a vis d'un groupe de jauge, qui n'est toutefois plus le groupe ab\'elien U(1), mais SU(3). De m\^eme que les photons sont les vecteurs de l'interaction \'electromagn\'etique, les gluons sont les vecteurs de l'interaction forte entre quarks. De cette interaction entre quarks r\'esulte une interaction forte entre nucl\'eons, qui est un peu \`a l'interaction entre quarks ce que l'interaction de van der Waals entre atomes est \`a l'interaction \'electromagn\'etique. Les quarks sont au nombre de 6, distingu\'es par 6 "saveurs", mais il suffit de 2 saveurs (baptis\'ees up et down, $u$ et $d$) pour fabriquer un proton de formule $uud$ et un neutron de formule $udd$. On retrouve ainsi les 2 \'etats de Heisenberg et l'invariance par $SU(2)$. En ajoutant le quark appel\'e {\it strange}, on peut obtenir d'autres particules dont la sym\'etrie est repr\'esent\'ee par la figure \ref{fig3}. 
Plus tard, en 1970, Glashow,  Iliopoulos
et  Maiani "invent\`erent" un quatri\`eme quark; et comme on le verra, ce n'\'etait pas fini.

\subsection{Le  mod\`ele standard et les   interactions faibles}
\label{ssec13'''}

Apr\`es avoir d\'ecrit les interactions fortes, qui assurent la coh\'esion des 
quarks dans les nucl\'eons et celle des nucl\'eons dans le noyau, les th\'eoriciens des particules
 les englob\`erent, avec l'\'electro-dynamique quantique, dans une th\'eorie plus unitaire, appel\'ee 
mod\`ele standard, et qui inclut aussi les interactions faibles. Le mod\`ele standard est une th\'eorie de jauge, 
fond\'ee sur un groupe de jauge $SU(3) \times  SU(2) \times U(1)$. Le groupe SU(3) sert, comme on l'a vu, 
\`a d\'ecrire les interactions fortes. On a vu que U(1) correspond \`a l'\'electromagn\'etisme. 
Est-ce \`a dire alors que SU(2) d\'ecrit les interactions faibles ? C'est plus compliqu\'e, 
car la sym\'etrie SU(2) X U(1) se brise spontan\'ement. Il n'en reste qu'une sym\'etrie U(1) 
distincte de la sym\'etrie U(1) d'o\`u on est parti, et pourvue de bosons de jauge de masse nulle, 
qui sont les photons. Mais les bosons de jauge li\'es \`a la sym\'etrie SU(2) ont acquis une masse non nulle. 
Ce sont les bosons de jauge de l'interaction faible (W+, W- et Z$_0$).

On attendrait pourtant de nouveaux bosons de masse nulle, ceux que le th\'eor\`eme de Goldstone associe \`a une sym\'etrie continue bris\'ee. Or ces bosons ne sont pas observ\'es. Ce paradoxe a \'et\'e r\'esolu par l'introduction d'un nouveau boson de masse non nulle, appel\'e boson de Higgs, qui r\'esulte d'une recombinaison du boson qui voudrait \^etre de Goldstone, lequel est ainsi phagocyt\'e et n'appara\^{\i}t pas. Ce "m\'ecanisme de Higgs" est propre aux sym\'etries de jauge. Dans la compr\'ehension de ce probl\`eme, le r\^ole de Nambu, prix Nobel 2008, fut essentiel. Il fut un pr\'ecurseur du boson de Goldstone, et fit beaucoup pour la compr\'ehension du m\'ecanisme de Higgs, que nous d\'ecrivons un peu plus en d\'etail dans l'appendice \ref{secA4}.

Quant au th\'eor\`eme de Goldstone, fabriqu\'e pour la physique des particules \'el\'ementaires, il n'y trouve gu\`ere d'applications exactes. Comme application approch\'ee on peut citer les pions, qui doivent leur faible masse au fait qu'ils sont approximativement des bosons de Goldstone li\'es \`a la violation d'une sym\'etrie approximative par rapport \`a la chiralit\'e.

\subsection{Au del\`a du  mod\`ele standard : supersym\'etrie, etc.}
\label{ssec13''''}

Le mod\`ele standard est probablement correct, en ce sens que toutes ses pr\'edictions sont 
v\'erifi\'ees exp\'erimentalement, sauf le boson de Higgs non encore observ\'e. Mais on aimerait avoir un mod\`ele qui pr\'edise davantage. Par exemple, le mod\`ele standard n'explique pas pourquoi la charge du proton est \'egale \`a celle de l'\'electron. Cette interrogation est parmi celle qui ont 
amen\'e \`a des th\'eories de grande unification, fond\'ees par exemple sur le groupe SU(5) ou 
sur SO(10). Le m\'ecanisme de Higgs y fonctionne \`a r\'ep\'etition et multiplie le nombre des 
particules \'el\'ementaires, que le mod\`ele standard avait r\'eussi \`a diminuer un peu. Puis on a envisag\'e une sym\'etrie entre bosons et fermions, bien entendu destin\'ee \`a \^etre bris\'ee : c'est la supersym\'etrie. Il restait \`a trouver une th\'eorie qui englobe la gravitation : les espoirs reposent actuellement sur la th\'eorie des cordes. Elle rompt avec le concept de particules ponctuelles, ce que l'on peut trouver satisfaisant ; il est vrai qu'il faut pour cela passer dans un espace \`a 10 dimensions. C'est le coeur, disait Pascal, qui sent qu'il y a trois dimensions 
dans l'espace; mais le coeur a ses raisons que la raison ne connait point, ajoutait-il. 
De toute fa\c con Einstein avait d\'ej\`a ajout\'e une dimension!

\subsection{Mati\`ere et antimati\`ere}
\label{ssec13'''''}

Disons quelques mots de la sym\'etrie entre mati\`ere et antimati\`ere, dont la rupture myst\'erieuse a justifi\'e 
le prix Nobel de Makoto Kobayashi et Toshihide Maskawa. Pendant des ann\'ees, 
les observateurs ont scrut\'e 
le ciel \`a la recherche d'anti-univers. En vain ; jusqu'en 1964 o\`u une exp\'erience sur le kaon r\'ev\'ela une violation des lois de la sym\'etrie. La formation de la mati\`ere et de l'antimati\`ere ne se fait pas via un processus \'equitable. Lors du Big Bang, la mati\`ere et l'antimati\`ere cr\'e\'ees se sont annihil\'ees. N'est rest\'e que cet exc\'edent de mati\`ere. Mais comment expliquer la violation de la sym\'etrie? C'est Makoto Kobayashi et Toshihide Maskawa qui, en 1972, d\'emontr\`erent (en exploitant une id\'ee du romain Nicola Cabibbo) que si la mati\`ere n'est pas 
constitu\'ee de quatre  quarks - comme on le croyait jusqu'alors - 
mais de six, alors la violation de la 
sym\'etrie se fait de fa\c con naturelle. Ils ajoutent ainsi \`a la liste des quarks existants (up, down, charme, 
\'etrange), les quarks bottom et top (aussi appel\'es v\'erit\'e et beaut\'e). Le premier
fut d\'ecouvert quelques ann\'ees plus tard, le second un peu plus tard encore.

Alors que la description des \'electrons et des photons par l'\'electrodynamique quantique 
est une th\'eorie unique et achev\'ee, il n'en est pas de m\^eme pour les autres particules. 
Pour les d\'ecrire, on peut h\'esiter entre diverses th\'eories dont aucune n'est totalement satisfaisante : les unes parce qu'elles n'expliquent pas tout, les autres parce qu'elles reposent sur des hypoth\`eses difficilement v\'erifiables. Toutes ces th\'eories font cependant grand usage de la notion de sym\'etrie et de la notion de groupe.

\section{Conclusion}
\label{sec14}

La sym\'etrie joue un grand r\^ole en physique. La brisure de sym\'etrie \'egalement. Elle s'introduit naturellement dans la th\'eorie des transitions de phase. Moins naturellement dans les th\'eories des particules \'el\'ementaires, o\`u elle est en partie le fruit de l'imagination fertile de th\'eoriciens en qu\^ete, l\'egitime, d'unification. 
Nous n'avons pas parl\'e d'autres domaines comme la m\'ecanique des fluides, o\`u la rupture de sym\'etrie est fr\'equemment observ\'ee. Des exemples sont l'instabilit\'e de Rayleigh-Plateau (fragmentation d'un jet cylindrique sous l'effet de la tension superficielle) ou l'instabilit\'e de Rayleigh-B\'enard (formation de rouleaux de convection dans une casserole chauff\'ee par en dessous).

Les groupes de transformations ne sont pas les seules structures de groupe qui apparaissent en physique. Les groupes de permutations sont un autre exemple. La statistique de Bose et la statistique de Fermi associent au groupe  des permutations de N particules identiques deux repr\'esentations irr\'eductibles diff\'erentes, mais si simples qu'il n'y a pas lieu d'invoquer la th\'eorie des groupes. Il n'en est pas de m\^eme si on cherche \`a d\'ecouvrir d'autres mod\`eles statistiques que ceux de Bose et de Fermi. Car on en a d\'ecouvert, mais \`a 2 dimensions seulement. Selon Wikipedia, le concept d'anyon peut servir pour d\'ecrire les couches de graph\`ene ou l'effet Hall quantique, ou pour les ordinateurs quantiques. Il convient aussi de mentionner le "groupe de renormalisation" qui permet un traitement math\'ematique 
pr\'ecis des transitions de phase. C'est bien un ensemble de transformations, mais plus abstraites que celles qui interviennent par exemple en cristallographie ; ces transformations sont plut\^ot des changements de variables dans un syst\`eme d'\'equations. D'autre part le groupe de renormalisation  n'est pas vraiment un groupe mais plut\^ot un semi-groupe, car les transformations ne sont pas inversibles, ou du moins on ne se pr\'eoccupe pas de leur trouver un inverse qui n'aurait gu\`ere d'int\'er\^et. 

Ainsi, la sym\'etrie est une notion partout pr\'esente dans la nature.
Pour l'exploiter quantitativement, la th\'eorie des groupes est  un outil puissant.
Elle requiert  des recettes qui sont efficaces quand on a appris \`a 
bien les manier, mais dont l'apprentissage n\'ecessite de la part de l'\'etudiant une certaine t\'enacit\'e. 

Ajoutons pour finir que  la th\'eorie des groupes est bien plus qu'un livre de recettes: elle \'eclaire 
souvent les lois physiques en 
les reliant \`a la simple g\'eom\'etrie.

\vskip1cm
Je remercie chaleureusement Jean Iliopoulos et Pierre Fayet pour des le\c cons \'el\'ementaires 
sur les particules du m\^eme nom, et pour avoir relu le manuscrit. Je suis reconnaissant \`a
Roger Balian, Manuel Houzet et Michail Zhitomirskii pour de pr\'ecieuses explications.






\appendix                     

\section{Structure \'electronique dans un r\'eseau hexagonal plan. Points de Dirac du graph\`ene}
\label{secA1}

\begin{figure}[t]
\centering
\includegraphics*[width=70mm, ]{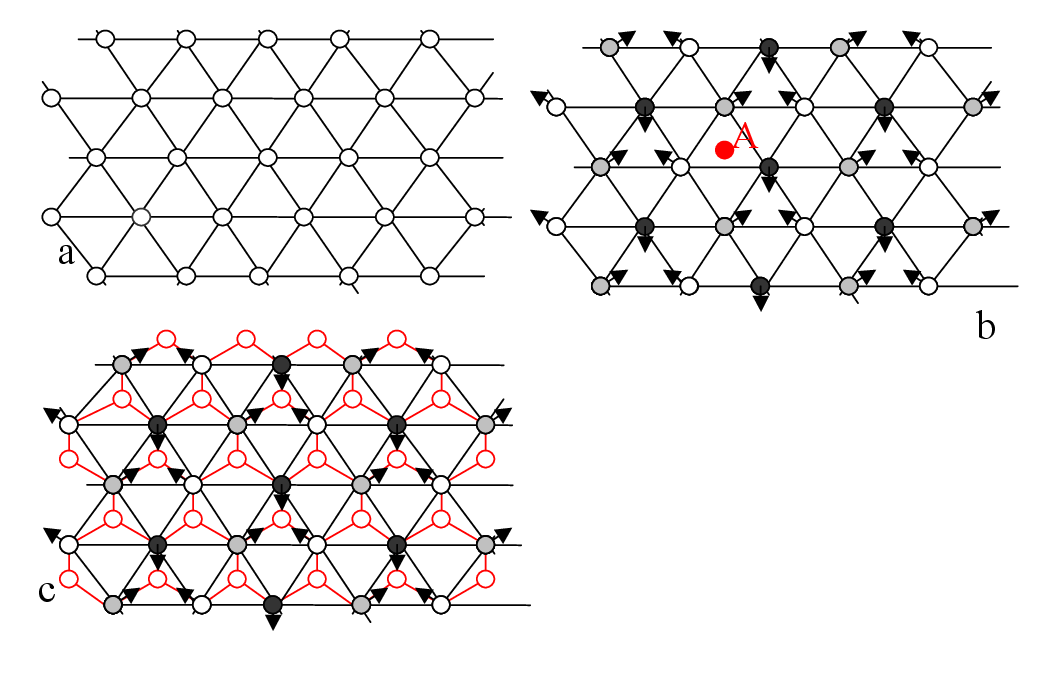}
\caption{ a) Un r\'eseau triangulaire plan. b) Une structure magn\'etique qui conserve la sym\'etrie de rotation d'ordre 3. c) En ajoutant un autre r\'eseau triangulaire on obtient un r\'eseau en nid d'abeille. En tout site du second r\'eseau, le champ mol\'eculaire d\^u au premier r\'eseau est nul.
}
\label{figA}
\end{figure}

Le graph\`ene est du graphite r\'eduit \`a une couche monoatomique. Sa structure de bande pr\'esente une particularit\'e remarquable que nous allons d\'emontrer. Dans l'approximation dite des "liaisons fortes", le hamiltonien d'un \'electron est 
\begin{equation}
{\mathcal H}=\sum_{ij}\gamma_{ij}c_i^+c_j
\label{A1}
\end{equation}
o\`u $i$ et $j$ d\'esignent les sites du r\'eseau qui dans le cas pr\'esent est un r\'eseau en nid d'abeilles, $c_i$ et $c_i^+$ d\'etruisent et cr\'eent un \'electron au site $i$, et les nombres $\gamma_{ij}$ ont les propri\'et\'es de sym\'etrie du r\'eseau. Les \'energies des \'etats \'electroniques sont les valeurs propres de la matrice $\Gamma $ des $\gamma_{ij}$. S'il n'y avait qu'un site par maille, elles seraient obtenues par une transformation de Fourier. Mais le r\'eseau en nid d'abeilles a 2 atomes par maille. Il est fait de 2 r\'eseaux de Bravais plans triangulaires 1 et 2 qui s'interp\'en\`etrent. Les composantes $u^i$ de tout vecteur propre s'\'ecrivent $u_i = u_1 \exp(i{\bf k}.{\bf R})$ si $i$ est le point 
{\bf R} du r\'eseau 1, et $u_i = u_2 \exp(i{\bf k}.{\bf R})$ si $i$ est le point 
{\bf R} du r\'eseau  2. Ceci peut s'\'ecrire $u_{R\alpha}  = u_\alpha  \exp(i{\bf k}.{\bf R})$ avec $i=1$ ou 2. Le nombres $u_1$ et $u_2$ sont vecteurs propres de la matrice
\begin{equation}
\Gamma({\bf k})   =\left[
\begin{array}{cc}
\gamma _{11}({\bf k}) & \gamma _{12}({\bf k})\\
\gamma _{21}({\bf k}) & \gamma _{22}({\bf k})
\end{array}
\right]
=\left[
\begin{array}{cc}
\gamma _{11}({\bf k}) & \gamma _{12}({\bf k})\\
\gamma _{12}^*({\bf k}) & \gamma _{11}({\bf k})
\end{array}
\right]
\label{A2}
\end{equation}
 o\`u
 $$
 \gamma_{\alpha \gamma }({\bf k})= \sum_{R',\gamma } \exp[i{\bf k}.({\bf r}_{R\alpha }-{\bf r}_{R'\gamma })]
 $$
Dans un premier temps, on va se d\'esint\'eresser du r\'eseau 2. A tout point ${\bf R} $ du r\'eseau 1 
(Fig. \ref{figA}a), on peut associer le scalaire complexe $\exp[i{\bf k}.({\bf R})$. Mais on peut aussi pr\'ef\'erer lui associer le vecteur bidimensionnel r\'eel 
\begin{equation}
{\bf v}({\bf R})=[\cos({\bf k}.{\bf R}), \sin({\bf k}.{\bf R})]	
\label{A2'}
\end{equation}
qui peut par exemple repr\'esenter un spin dans un probl\`eme de magn\'etisme. dont les moments magn\'etiques seraient localis\'es aux points R d'un r\'eseau triangulaire (Fig. \ref{figA}a). Il est naturel de se demander s'il existe une structure magn\'etique qui conserve la sym\'etrie de rotation d'ordre 3. La r\'eponse est oui, comme le montre la Figure \ref{figA}b. Si O est le centre de l'un des triangles, la structure 4b est invariante par une rotation du r\'eseau de 2$\pi $/3 autour de O, accompagn\'ee d'une rotation des vecteurs (spins) de 2$\pi $/3. Or 
l'invariance par rotation entra\^{\i}ne que le "champ mol\'eculaire" au point O est nul, quels que soient 
les $\gamma _{ij}$. 
 \begin{equation}
\sum_R \gamma_{OR}{\bf v}({\bf R})=0	
\label{A3}
\end{equation}
o\`u ${\bf R}$ d\'esigne les sites du r\'eseau 1. Or O est un site du r\'eseau 2, comme on le voir par la figure \ref{figA}c. 
Les composantes des "spins" de la Figure \ref{figA}b sont les parties r\'eelle et imaginaire de 
 \begin{equation}
\psi _K({\bf R})= \exp(i{\bf K}.{\bf R})
\label{A4}
\end{equation}
o\`u 
 \begin{equation}
K_x=\frac{2\pi}{3b} \;\;\;\;;\;\;\;\;\;\;K_y=\frac{2\pi}{b\sqrt {3}}, 
\label{A5}
\end{equation}
et $b $ est la distance entre les sites du r\'eseau triangulaire. Soit en fonction de la distance interatomique $a=b/\sqrt{3}$  du r\'eseau en nid d'abeille, 
\begin{equation}
K_x=\frac{2\pi}{3b\sqrt {3}} \;\;\;\;;\;\;\;\;\;\;K_y=\frac{2\pi}{3b}, 
\label{A6}
\end{equation}

La relation (\ref{A3}) s'\'ecrit donc 
\begin{equation}
\gamma _{12}({\bf K})=0
\label{A6'}
\end{equation}

La matrice $\Gamma({\bf K})$ est donc d\'eg\'en\'er\'ee, avec ses 2 valeurs propres \'egales. Revenant au probl\`eme \'electronique, on voit qu'il existe un \'etat \'electronique doublement d\'eg\'en\'er\'e, quelle que soit la port\'ee des $\gamma _{ij}$. C'est en fait une tr\`es bonne approximation de supposer (comme on le fait g\'en\'eralement) que 
$\gamma _{ij}$ n'est non nul que pour des sites i et j voisins (appartenant donc l'un au r\'eseau 1, l'autre au r\'eseau 2). Dans ce cas on peut montrer que si ${\bf k}$ n'est pas \'egal \`a ${\bf K}$, l'un des \'etats \'electroniques a une \'energie $\epsilon ^-({\bf k})<\epsilon ({\bf K})$ alors que l'autre \'energie est 
$\epsilon ^+({\bf k})>\epsilon ({\bf K})$. Comme dans le graph\`ene pur il y a exactement un \'electron par site, on a donc \`a temp\'erature nulle une bande pleine et une bande vide qui se touchent en un nombre discret de points de l'espace r\'eciproque, avec un gap nul. Ces points sont appel\'es points de Dirac.

\section{Recette pour obtenir les vecteurs propres d'une matrice qui appartiennent \`a une repr\'esentation irr\'eductible particuli\`ere}
\label{secA1b}

Nous voulons diagonaliser une matrice carr\'ee  $M$ (\`a $n\times n$ \'el\'ements).
Cette matrice  agit sur des matrices colonnes ou vecteurs \`a $n$ \'el\'ements que nous  noterons $ \ket{\varphi} $
et qui constituent un espace vectoriel $ \mathcal {E} $.  
Par hypoth\`ese, $M$ est invariante par les op\'erations $g$ d'un groupe $G$.  
Ceci veut dire que $O(g)MO(g)^{-1}=M$, o\`u $O(g)$ est une matrice qui d\'ecrit l'action de $g$ sur les vecteurs $ \ket{\varphi} $. Les matrices $O(g)$ forment une repr\'esentation de $G$, que nous supposerons r\'eductible. Si elle
ne l'est pas, la th\'eorie des groupes ne peut apporter aucune aide.

Le groupe $G$ admet des repr\'esentations irr\'eductibles par des matrices $ \mathcal{D}^{(p)}(g)$ de dimensions $d_p$, dont 
les \'el\'ements sont $ \mathcal{D}^{(p)}_{\lambda \mu }(g)$.  

On d\'emontre que les matrices $ \mathcal{D}^{(p)}(g)$ v\'erifient alors la propri\'et\'e 
d'orthogonalit\'e (Streitwolf 1971, Canals 2009)\footnote{Sacrifiant la 
rigueur pour la simplicit\'e, nous omettons certains d\'etails importants. Notamment,
les  matrices $ \mathcal{D}^{(p)}(g)$ sont suppos\'ees unitaires. Cela est possible parce que toute 
repr\'esentation est \'equivalente \`a une  repr\'esentation par des matrices unitaires
 (Streitwolf 1971, Canals 2009).}

\begin{equation}
 \sum_g \mathcal{D}^{(p)}_{\lambda \mu }(g) \mathcal{D}^{(p')}_{\lambda ' \mu '}(g) = (n/d_p) 
\delta_{pp'} \delta_{\lambda \lambda '}\delta_{\mu \mu '}
\label{App1}
\end{equation}

On dit qu'un vecteur $ \ket{e_\lambda } $ de $ \mathcal {E} $ appartient \`a la repr\'esentation irr\'eductible $p$ si 

\begin{equation}
O(g) \ket{e_\lambda }= \sum_\mu  \mathcal{D}^{(p)}_{\lambda \mu }(g) \ket{e_\mu }
\label{App2}
\end{equation}

D'apr\`es le th\'eor\`eme du d\'eveloppement (expansion theorem, Streitwolf 1971) tout vecteur  $ \ket{\varphi} $
peut s'\'ecrire comme une somme de
 composantes dont chacune  appartient une repr\'esentation irr\'eductible $p$. 

\begin{equation}
\ket{\varphi} = \sum _p \sum _\lambda  \sum _\alpha  \ket{e^{(p)}_{\lambda \alpha  }}
\label{App3}
\end{equation}
o\`u $ 1 \leq \lambda \leq d_p $, $p$ d\'esigne les diverses repr\'esentations  irr\'eductibles, et on a introduit un indice 
suppl\'ementaire $ \alpha $ qui prend les valeurs enti\`eres jusqu'\`a une valeur $q_p$ qui d\'epend de la matrice $M$, 
et peut \^etre nul pour certaines valeurs de $p$. Il existe une recette pour d\'eterminer $q_p$
(Streitwolf 1971, Canals 2009), mais nous ne l'utiliserons pas. Les  vecteurs 
$ \ket{e^{(p)}_{\lambda \alpha  }}$ seront suppos\'es orthonorm\'es. Si ces  vecteurs 
\'etaient pris comme vecteurs de base, 
la matrice $M$ aurait la forme diagonale par blocs
(\ref{reduc22}). Le probl\`eme est que les vecteurs $ \ket{e^{(p)}_{\lambda \alpha  }}$
ne sont pas connus. On va voir comment on peut quand m\^eme mettre 
la matrice $M$ sous la forme diagonale par blocs.

Pour chaque vecteur  $\ket{\varphi}$ nous voulons d\'eterminer chaque terme de la somme  (\ref{App3}). En faisant cela 
pour $n$ vecteurs  $\ket{\varphi}$ ind\'ependants, nous 
obtiendrons, pour chaque $p$, des  vecteurs qui sont des combinaisons lim\'eaires des
$ \ket{e^{(p)}_{\lambda \alpha  }}$. Leur orthogonalisation fournit $d_p q_p$ vecteurs ind\'ependants
qui sous-tendent un espace  $ \mathcal {E}_p $. 
En faisant cela pour toutes les valeurs de $p$, on obtient une base dans laquelle 
la matrice $M$ est diagonale par blocs.

Apr\`es avoir \'enonc\'e le principe de la m\'ethode, donnons la recette. Si on multiplie (\ref{App3}) 
par $[\mathcal{D}^{(p)}_{\lambda \mu }]^* (g) O(g)$ et si on somme sur $g$, on trouve en utilisant 
(\ref{App1}) et (\ref{App2})

\begin{equation}
\sum _g [\mathcal{D}^{(p)}_{\lambda \mu }]^\star(g) O(g) \ket{\varphi} = 
(n/d_p) \sum _\alpha \varphi ^{(p)} _{\lambda \alpha } \ket{e^{(p)}_{\mu \alpha  }}
\label{App4}
\end{equation}

Si maintenant on fait $ \lambda =\mu $ et qu'on somme sur $\lambda $, le second membre se r\'eduit \`a
 un facteur pr\`es \`a la projection de (\ref{App3}) sur l'espace  $ \mathcal {E}_p $:

\begin{equation}
\sum _g \sum _\lambda [\mathcal{D}^{(p)}_{\lambda \mu }]^\star(g) O(g) \ket{\varphi} = 
n P_p \ket{\varphi}
\label{App5}
\end{equation}
o\`u 
\begin{equation}
P_p = (1/n) \sum _g \sum _\lambda [\mathcal{D}^{(p)}_{\lambda \mu }]^\star(g) O(g) 
\label{App6}
\end{equation}
 est l'op\'erateur de projection sur  $ \mathcal {E}_p $. Il n'y a plus 
qu'\`a achever le programme en faisant agir $P_p$ sur $n$ vecteurs ind\'ependants, orthogonaliser les
vecteurs obtenus, et r\'ep\'eter l'op\'eration pour toutes les repr\'esentations irr\'eductibles $p$.
Dans la base ainsi obtenue, la matrice $M$ est diagonale par blocs.



\section{Repr\'esentations spinorielles du groupe de Lorentz  et spineurs de Dirac}
\label{Cartan}

{\it Repr\'esentation spinorielle bidimensionnelle irr\'eductible du groupe de Lorentz}

Pour l'obtenir, il convient d'associer \`a tout quadrivecteur ${\bf x}=(ct,x,y,z)$ 
la matrice bidimensionnelle  hermitique 

 \begin{equation}
X = \left[
\begin{array}{cc}
ct+z & x+iy \\
 x-iy & ct-z
\end{array}
\right] = t + x\sigma _x+ y\sigma _y+ z\sigma _z 
\label{Cartan1}
\end{equation}

On voit que $\det X = c^2t^2-x^2-y^2-z^2$ est la norme de Minkowski de ${\bf x}$. Elle doit \^etre conserv\'ee dans une transformation de Lorentz.

Nous allons montrer que les matrices $2\times 2$ complexes $B$ de d\'eterminant 1 
permettent d'obtenir une repr\'esentation spinorielle 
du groupe de Lorentz.
Dans ce but, nous  consid\'erons la matrice.
 \begin{equation}
X'=B X B^+
\label{Cartan2}
\end{equation}
o\`u $B^+$ d\'esigne la matrice hermitique conjugu\'ee de $B$. La matrice $X'$ est hermitique et son d\'eterminant est \'egal \`a $\det X$.
Il d\'efinit donc un quadrivecteur ${\bf x}'$ par la formule (\ref{Cartan1}). Celle-ci est donc une
transformation de Lorentz. 

Par exemple une  translation uniforme dans la direction $z$ correspond \`a $B= \exp(-\varphi \sigma _z /2)$. En effet,
en utilisant les propri\'et\'es d'anticommutation des matrices de Pauli, on voit que

$$
BXB^+ = BXB  = \exp(-\varphi \sigma _z /2)( ct + x\sigma _x+ y\sigma _y+ z\sigma _z )
\exp(-\varphi \sigma _z /2)
$$
$$
 = ( ct +  z\sigma _z )\exp(-\varphi \sigma _z) 
+ ( x\sigma _x+ y\sigma _y )
$$

Ceci est la transformation de Lorentz (\ref{Lorentz}), qui correspond  \`a une 
translation uniforme dans la direction $z$,
\`a vitesse uniforme $v=\beta c$ avec $\sinh \varphi =1/\sqrt{1-\beta ^2}$.

Un calcul analogue montre qu'une rotation des axes  $x$ et $y$ 
d'angle $\varphi $ autour de l'axe $z$  correspond \`a 
$B= \exp(-i\varphi \sigma _z /2)$ comme dans le cas non relativiste. 

Les r\`egles sont analogues pour les rotations autour de $x$ and $y$ et les  translations uniformes le long de $x$ et $y$. Ceci montre que toute 
transformation de Lorentz correspond \`a une matrice B ... ou plut\^ot  \`a deux matrices  puisque $B$ peut 
\^etre remplac\'ee par $-B$ dans (\ref{Cartan2}).

Ainsi, dans la  repr\'esentation spinorielle du  groupe de Lorentz que nous venons de d\'efinir,  
l'op\'erateur $\sigma _\alpha $ d\'ecrit \`a la fois 
une rotation infinit\'esimale autour de l'axe $\alpha $  et une translation  le long de cet axe, la 
seule diff\'erence \'etant le facteur $i$. Comme ces deux op\'erations sont physiquement tr\`es diff\'erentes, la
repr\'esentation spinorielle bidimensionnelle est peu appropri\'ee \`a une description quantique relativiste.

{\it Repr\'esentation de Dirac du groupe de Lorentz.}

Si des matrices $B(g)$ forment une repr\'esentation d'un groupe non commutatif  $G$ dont les \'el\'ements 
sont $g$, les matrices hermitiques conjugu\'ees $B^+(g)$ ne  forment  g\'en\'eralement pas une repr\'esentation, car  $B^+(g)B^+(g')=B^+(g'g)$
est g\'en\'eralement diff\'erent de $B^+(gg')$. Par contre les matrices 
$[B^+(g)]^{-1}$  forment bien une repr\'esentation.
Si $G$ est le groupe de Lorentz et $B(g)$ sont les matrices d\'efinies pr\'ec\'edemment, 
ces deux repr\'esentations ne sont pas \'equivalentes.
En effet, l'\'equivalence de deux repr\'esentations $B$ et $B'$ implique $B'(g)=UB(g)U^{-1}$ et par suite Tr $B$= Tr $B'$. Cette relation n'est pas v\'erifi\'ee, par exemple pour une matrice diagonale, $B'_{11}+B'_{22}= (B_{11}^*+B_{22}^*)/(B_{11}B_{22})^*=B_{11}^*+B_{22}^*$ 
n'est g\'en\'eralement pas \'egal \`a $B_{11}+B_{22}$. 

La repr\'esentation de Dirac associe \`a toute  transformation de Lorentz $g$ la matrice $4 \times 4$ 

 \begin{equation}
 \left[
\begin{array}{cc}
B(g) & 0 \\
 0 & [B^+(g)]^{-1}
\end{array}
\right] 
\label{Cartan3}
\end{equation}

Cette repr\'esentation r\'eductible op\`ere sur un espace de spineurs \`a 4 dimensions  et la fonction d'onde qui
ob\'eit \`a l' \'equation  de Dirac est un \'el\'ement  de cet espace.
Elle est form\'ee de 2 spineurs \`a 2 dimensions, dont l'un se  transforme par $B$, l'autre par $(B^+)^{-1}$ 
quand on fait une transformation de Lorentz.

\section{Invariance de jauge en \'electrodynamique}
\label{secA2}

Nous nous limiterons au cas d'un \'electron en interaction avec le champ \'electromagn\'etique. L'\'electron (de masse 
$m$ et de charge $q=-e$) est repr\'esent\'e par une fonction d'onde \`a 4 composantes (celle de l'\'equation de Dirac) et le champ \'electromagn\'etique par le potentiel vecteur 
\`a 4 composantes $A_\mu$. Jauch \& Rohrlich (1955) 
\'ecrivent la fonction suivante qu'ils appellent ``lagrangien'' 
(mais qui pour d'autres auteurs est la densit\'e de lagrangien) :
\begin{equation}
{\mathcal L}=-(1/2) \partial_\mu A_\nu \partial^\mu A^\nu -{\bar \Psi} (\gamma_\mu\partial^\mu+m)\Psi 
-ie{\bar \Psi}\gamma_\mu A ^\mu \Psi
\label{A7}
\end{equation}
o\`u $\gamma_\mu$ d\'esigne les 4 matrices $4\times 4$ d\'efinies plus haut et la  sommation 
sur les indices est sous-entendue (convention d'Einstein).

Le lagrangien (A1) permet de d\'efinir l'action $S=\int {\mathcal L} d^4 x$, o\`u l'int\'egrale est sur le domaine compris entre 2 hyperplans de l'espace-temps relativiste. Testa (1993) remarque que l'action (\ref{A7}) est invariante par la transformation (\ref{electro2}) combin\'ee \`a la transformation de  la fonction d'onde \'electronique $ \Psi ({\bf x})$ en
\begin{equation}
\Psi'({\bf x})= \exp [i\Lambda({\bf x}) \Psi ({\bf x})
\label{A8}
\end{equation}

L'\'equation de Dirac n'est pas facile \`a g\'en\'eraliser \`a plus d'un \'electron. Il est donc int\'eressant de 
d\'emontrer (\ref{A8}) \`a partir de l'\'equation non relativiste de Schr\"odinger. Cela est fait par 
Cohen-Tannoudji et al. (1987), pages 169 et 170.

\section{Un formalisme pour les interactions fortes}
\label{secA3}

Pour voir \`a quoi ressemblent les \'equations des particules \'el\'ementaires en \'evitant des complications excessives, 
nous nous bornerons aux interactions nucl\'eaires fortes, d\'ecrites par la chromodynamique quantique.
Ce formalisme reposait sur un lagrangien que nous \'ecrirons d'une part pour mesurer 
l'analogie et la diff\'erence avec l'\'electrodynamique quantique de l'appendice \ref{secA2}, d'autre part pour compter les variables et les param\`etres dont il d\'epend:
 \begin{equation}
{\mathcal L}_{QCD}= 
{\bar \Psi}_{ip}(i\gamma^\mu\partial_\mu-m_p)\Psi_{ip}-g G_\mu ^a 
{\bar \Psi}_{ip} \gamma^\mu T_{ij}^a \Psi_{jp}-(1/4) G_{\mu \nu} ^a G^{\mu \nu} _a
\label{A11}
\end{equation}

Dans cette formule, les spineurs \`a 4 composantes $\Psi_{ip}$ d\'esignent les champs de quarks; 
l'indice $p$=1,2,... 6 correspond \`a la "saveur", et l'indice $i$=1, 2 ou 3 correspond \`a la "couleur". Le champ $G$ est celui des "gluons". Comme les photons, les gluons sont de masse nulle. Les quantit\'es $T^a_{ij}$ 
sont donn\'ees par la th\'eorie des groupes (matrices de Gell-Mann). La constante de couplage $g$ est un param\`etre, qui s'ajoute aux 6 masses. Au total il y a 7 param\`etres. Que devient la sym\'etrie dans tout \c ca ? On pourrait perdre espoir en sachant que les masses des 6 quarks sont diff\'erentes. Toutefois, celles des quarks $u$ et $d$ sont de l'ordre du MeV (\`a un facteur $c^2$ pr\`es), c'est-\`a-dire faibles par rapport aux masses des nucl\'eons qu'ils composent, lesquelles sont de l'ordre du GeV. Il n'y a donc pas l\`a de quoi d\'et\'eriorer outre mesure la sym\'etrie entre proton et neutron (en ignorant, bien entendu, l'\'electromagn\'etisme). Les choses se g\^atent quand on s'int\'eresse aux autres quarks qui sont nettement plus lourds, de quelques centaines de MeV \`a quelques centaines de GeV.

\section{Interactions \`a longue port\'ee et m\'ecanisme de Higgs}
\label{secA4}

On peut se demander ce que devient une onde sonore dans un gaz quand les particules deviennent charg\'ees. L'onde en question a une fr\'equence non nulle pour $q=0$; ce n'est plus en fait une onde sonore... C'est un plasmon, 
et le gaz charg\'e est un plasma. Esquissons la d\'emonstration. L'\'equation du mouvement est 
$d^2u_q/dt^2=-V_q q^2u_q$, 
o\`u $u_q$ est la transform\'ee de Fourier du d\'eplacement et $V_q$ est la transform\'ee de Fourier du potentiel. 
Pour des interactions \`a courte port\'ee, $V_q$ est fini pour q=0, et la fr\'equence $\omega _q$ est 
proportionnelle \`a $q$. Pour des interactions coulombiennes, $V_q$  est proportionnel \`a $1/q^2$ et la fr\'equence de plasma est non nulle pour q=0. Cette remarque constituait le d\'ebut d'un article qu'Anderson publia en 1963. L'auteur consid\'erait aussi le cas analogue de la supraconductivit\'e. Le mode de Goldstone serait alors une oscillation de courant, 
$j(x,t)=j_0 \cos(qx-\omega t)$. Mais une telle oscillation dans l'espace impliquerait des amas de charge, dont l'\'energie serait finie \`a cause de la longue port\'ee de l'interaction coulombienne. Le mode de Goldstone de la supraconductivit\'e n'annule pas sa fr\'equence pour $q=0$. Dans le langage des  hautes \'energies, il est "massif" ; ce n'est plus en fait un mode de Goldstone.

Avec une grande perspicacit\'e, Anderson voyait une analogie entre l'absence exp\'erimentale de mode de fr\'equence nulle dans un plasma, et l'absence de mode de Goldstone dans des probl\`emes de physique des particules \'el\'ementaires o\`u il \'etait, \`a l'\'epoque, attendu, malgr\'e des r\'eflexions 
ant\'erieures de Schwinger, qui avait d\'ej\`a pressenti la solution, c'est-\`a-dire le boson de Higgs.  




\vskip4cm

\begin{thebibliography}{99}
\bibitem{Amara}  M. Amara (2009)  \'Ecole sur la sym\'etrie en Physique de la mati\`ere condens\'ee. 
A para\^{i}tre au Journal de Physique 4.
\bibitem{Anderson} 
P.W. Anderson, Phys. Rev. {\bf 130} (1963) 439.
\bibitem{Aroyo} M. I. Aroyo (2009)  \'Ecole sur la sym\'etrie en Physique de la mati\`ere condens\'ee. 
A para\^{i}tre au Journal de Physique 4.
\bibitem{Axel} F. Axel, S. Aubry (1981)  J. Phys. C: Solid State Phys. {\bf 14} 5433
\bibitem{Bertaut} E.F. Bertaut (1971) Journal de Physique Colloque C1, 32, 462 
\bibitem{Bouree}  F. Bour\'ee (2009)  \'Ecole sur la sym\'etrie en Physique de la mati\`ere condens\'ee. 
A para\^{i}tre au Journal de Physique 4.
\bibitem{Brout}R. Brout, F. Englert (2007) C. R. Physique {\bf 8}, 973
\bibitem{Byers}N. Byers (1996) http://fr.arxiv.org/abs/physics/9807044
\bibitem{Canals}  B. Canals (2009)  \'Ecole sur la sym\'etrie en Physique de la mati\`ere condens\'ee. 
A para\^{i}tre au Journal de Physique 4.
\bibitem{Cartan1913} E. Cartan (1913) Bull. Soc. Math. France {\bf 41}, 56.
\bibitem{Cartan} E. Cartan (1938) Le\c cons sur la th\'eorie des spineurs. Paris, H. Hermann.
\bibitem{CCT} C. Cohen-Tannoudji, J. Dupont-Roc, G. Grynberg (1987) Photons et atomes 
(EDP Sciences, CNRS editions, Les Ulis).
\bibitem{Comtet} A. Comtet (2005) {\it L'\'equation de Dirac} Lecture at the Universit\'e Piere et Marie Curie. http://cel.archives-ouvertes.fr/cel-00092970/fr/
\bibitem{Curie} P. Curie, Bulletin de la Soci\'et\'e min\'eralogique de France {\bf 7}, 418 (1884) et {\it Oeuvres de Pierre Curie}, available through Gallica (http://gallica.bnf.fr/)p. 79
\bibitem{Eckold}  G. Eckold (2009)  \'Ecole sur la sym\'etrie en Physique de la mati\`ere condens\'ee. 
A para\^{i}tre au Journal de Physique 4.
\bibitem{Englert} F. Englert, R. Brout, Phys. Rev. Lett. {\bf 13} (1964) 321.
\bibitem{GIM} S. L. Glashow, J. Iliopoulos, L. Maiani (1970)
Phys. Rev. D 2, 1285.
\bibitem{Goldstone61}J. Goldstone, Nuovo Cimento {\bf 19} (1961) 154.
\bibitem{Goldstone62}J. Goldstone, A. Salam, S. Weinberg, Phys. Rev. {\bf 127} (1962) 965.
\bibitem{Grenier} B. Grenier (2009). \'Ecole sur la sym\'etrie en Physique de la mati\`ere condens\'ee. 
A para\^{i}tre au Journal de Physique 4. 
\bibitem{Hamermesh}M. Hamermesh (1962) Group theory and its application to physical problems (Dover)
\bibitem{Higgs1}P.W. Higgs, Phys. Lett. {\bf 12} (1964) 132.
\bibitem{Higgs2}P.W. Higgs, Phys. Rev. Lett. {\bf 13} (1964) 508.
\bibitem{Higgs07}P. Higgs (2007) C. R. Physique {\bf 8}, 970
\bibitem{Houzet}  M. Houzet (2009)  \'Ecole sur la sym\'etrie en Physique de la mati\`ere condens\'ee. 
A para\^{i}tre au Journal de Physique 4.
\bibitem{Jauch} J.M. Jauch, F. Rohrlich (1955) The theory of photons and electrons (Addison-Wesley, Cambridge, U.S.A.)
P.H.E. Meijer, E. Bauer (2004) Group Theory: The Application To Quantum Mechanics 
(Dover).
\bibitem{Kleinert} H. Kleinert (1989)   {\it Gauge fields in condensed matter} (World scientific, Singapore).
\bibitem{Kreisel} J. Kreisel (2009)  \'Ecole sur la sym\'etrie en Physique de la mati\`ere condens\'ee. 
A para\^{i}tre au Journal de Physique 4.
\bibitem{Monroe} Monroe, Don, , New Scientist, 1 October 2008 
\bibitem{Noether} E. Noether, {\it Invariante Variationsprobleme}, Nachr. d. K\"onig. Gesellsch. d. Wiss. zu G\"ottingen, Math-phys. Klasse (1918), 235-257; English translation M. A. Travel, Transport Theory and Statistical Physics 1(3) 1971,183-207. 
\bibitem{Carvajal}  J. Rodriguez-Carvajal (2009). This school. 
\bibitem{Schweizer05}
J. Schweizer (2005) C.R. Physique 6, 375 and Corrigendum (2006) C.R. Physique 7, 823
\bibitem{Schweizer07} J. Schweizer, J. Villain, A.B. Harris (2007) Eur. Phys. J. Appl. Phys. 38, 41 
\bibitem{Schweizer09} J. Schweizer (2009). \'Ecole sur la sym\'etrie en Physique de la mati\`ere condens\'ee. 
A para\^{i}tre au Journal de Physique 4.
\bibitem{Streitwolf}H.W. Streitwolf (1971) Group theory in solid-state physics (Macdonalds, London).
\bibitem{Testa}M. Testa (1993) Articles `gauge, teorie di' and `spinore' in Enciclopedia delle Scienze fisiche, tome 2, p. 842 (Rome)
\bibitem{Toledano} P. Toledano (2009) \'Ecole sur la sym\'etrie en Physique de la mati\`ere condens\'ee. 
A para\^{i}tre au Journal de Physique 4.
\bibitem{Villain} J. Villain, M. Gordon (1980) J . Phys. C: Solid State Phys. {\bf 13} 3117
\bibitem{Wilczek}F. Wilczek (1982) Phys. Rev. Lett. {\bf 49}, 957
\bibitem{Wunsch} B. Wunsch, F. Guinea, F. Sols, New Journal of Physics {\bf 10} (2008) 103027
\bibitem{Yukawa} H. Yukawa (1949)  Nobel lecture. http://nobelprize.org/
\end{thebibliography}
\end{document}